\documentclass[12pt,a4paper]{article}
\usepackage[a4paper,top=23mm,bottom=23mm,left=25mm,right=25mm]{geometry}
\usepackage{amsmath,amssymb,graphicx,booktabs,array,caption,natbib,placeins,float}
\usepackage{ragged2e}
\usepackage{etoolbox}
\usepackage{xcolor}
\definecolor{refblue}{HTML}{1F5FAE}
\usepackage[unicode,colorlinks=true,linkcolor=refblue,citecolor=refblue,urlcolor=refblue]{hyperref}
\newcommand{\secref}[1]{\hyperref[#1]{Section~\ref*{#1}}}
\newcommand{\tabref}[1]{\hyperref[#1]{Table~\ref*{#1}}}
\newcommand{\figref}[1]{\hyperref[#1]{Figure~\ref*{#1}}}
\newcommand{\appref}[1]{\hyperref[#1]{Appendix~\ref*{#1}}}
\newcommand{\apptabref}[1]{\hyperref[#1]{Appendix Table~\ref*{#1}}}
\newcommand{\appfigref}[1]{\hyperref[#1]{Appendix Figure~\ref*{#1}}}
\newcommand{\eqnref}[1]{\hyperref[#1]{Equation~(\ref*{#1})}}
\DeclareRobustCommand{\sym}[1]{\textsuperscript{#1}}
\newcolumntype{C}[1]{>{\begin{minipage}[c]{#1}\centering\arraybackslash\setlength{\parskip}{0pt}\strut\ignorespaces}c<{\unskip\strut\end{minipage}}}
\newcolumntype{L}[1]{>{\begin{minipage}[c]{#1}\raggedright\arraybackslash\setlength{\parskip}{0pt}\strut\ignorespaces}c<{\unskip\strut\end{minipage}}}
\newcommand{\grouphead}[1]{\ensuremath{\vcenter{\hbox{\shortstack[c]{#1}}}}}
\newcommand{\tabnum}[1]{\mbox{\ensuremath{#1}}}
\newcommand{\regcoef}[3]{\mbox{\ensuremath{#1.#2}\if\relax\detokenize{#3}\relax\else\sym{#3}\fi}}
\newcommand{\regse}[2]{\tabnum{(#1.#2)}}
\newcommand{\regpair}[2]{\ensuremath{\vcenter{\hbox{\shortstack[c]{#1\\#2}}}}}
\newsavebox{\rowlabelbox}
\newcommand{\rowlabel}[1]{%
  \sbox{\rowlabelbox}{#1}%
  \ifdim\wd\rowlabelbox>\linewidth
    \ifdim\wd\rowlabelbox>1.12\linewidth #1\else\resizebox{\linewidth}{!}{\usebox{\rowlabelbox}}\fi
  \else\usebox{\rowlabelbox}\fi}
\newcommand{\regstat}[2]{\tabnum{#1.#2}}
\newcommand{\descnum}[2]{\tabnum{#1.#2}}
\newcommand{\incrnote}{Incremental within-$R^2$ is the difference in within-$R^2$ between the full and restricted models, expressed in percentage points (denoted by \% in tables and figures).}
\newcommand{\regnote}{Standard errors are two-way clustered by stock and calendar week and are reported in parentheses unless shown separately. \sym{***}, \sym{**}, and \sym{*} denote significance at the 1\%, 5\%, and 10\% levels, respectively.}
\newenvironment{exhibitnotes}{%
  \par\smallskip
  \justifying
  \setlength{\parindent}{0pt}%
  \setlength{\parskip}{0pt}%
  \linespread{1.0}\fontsize{8.5}{10.5}\selectfont
  \noindent\textit{Notes:}\enspace\ignorespaces
}{\par}
\AtBeginEnvironment{tabular}{\setlength{\aboverulesep}{1.1pt}\setlength{\belowrulesep}{1.1pt}\setlength{\cmidrulesep}{1.1pt}}
\usepackage{titlesec}
\DeclareFontFamily{OT1}{cmheading}{}
\DeclareFontShape{OT1}{cmheading}{bx}{n}{<-> cmbx12}{}
\newcommand{\headingfont}[2]{\fontencoding{OT1}\fontfamily{cmheading}\fontseries{bx}\fontshape{n}\fontsize{#1}{#2}\selectfont}
\newcommand{\papertitlefont}{\headingfont{16}{19.2}}
\newcommand{\sectionheadingfont}{\headingfont{14}{16.8}}
\newcommand{\subsectionheadingfont}{\headingfont{13}{15.6}}
\newcommand{\appendixsectionfont}{\headingfont{14}{16.8}}
\newcommand{\appendixsubsectionfont}{\headingfont{13}{15.6}}
\titleformat{\section}{\sectionheadingfont\raggedright}{\thesection.}{.5em}{}
\titlespacing*{\section}{0pt}{15pt}{8pt}
\titleformat{\subsection}{\subsectionheadingfont\raggedright}{\thesubsection}{.5em}{}
\titlespacing*{\subsection}{0pt}{11pt}{5pt}
\begin{document}
\begingroup
\renewcommand{\thefootnote}{\fnsymbol{footnote}}
\begin{center}
{\papertitlefont Price Discovery around Stock Market Reopenings:\\[5pt]
Evidence from 24/7 Equity Perpetuals\footnote{This draft is preliminary. Xiaoxing Liu acknowledges financial support from the Major Program of the National Social Science Fund of China (Grant No. 24\&ZD117). All errors are our own.}\par}
\vspace{8pt}
{\small Zhiyi Wang\textsuperscript{1,3}\qquad
Xiaoxing Liu\textsuperscript{1,2,3}\footnote{Corresponding author. Email address: wangzhiyi924@gmail.com.}\par}
\vspace{5pt}
{\setlength{\parskip}{2pt}\linespread{1.0}\fontsize{9}{11}\selectfont
\textsuperscript{1}\textit{School of Cyber Science and Engineering, Southeast University, 2 Southeast University Road, Jiangning District, Nanjing 211189, China}\par
\textsuperscript{2}\textit{School of Economics and Management, Southeast University, 2 Southeast University Road, Jiangning District, Nanjing 211189, China}\par
\textsuperscript{3}\textit{Laboratory of Philosophy and Social Sciences at Universities in Jiangsu Province-Fintech and Big Data Laboratory of Southeast University, 2 Southeast University Road, Jiangning District, Nanjing 211189, China}\par}
\end{center}
\endgroup
\setcounter{footnote}{0}

\begin{abstract}

Using transaction data on 24/7 equity perpetuals covering 421 U.S.-listed stocks across 22 venues, we examine the information in weekend prices and their incremental value as stock trading resumes. In the baseline sample, a one-percentage-point higher weekend perpetual return is associated with a 59.2-basis-point higher cumulative stock return at the regular-session open. Controlling for cumulative stock returns five minutes into overnight trading sharply reduces perpetual returns' incremental explanatory power. Conditional on cumulative stock returns five minutes before the regular-session open, the remaining opening effect passes an equivalence test with bounds of $\pm10$ basis points per percentage point of perpetual return.

\end{abstract}

\section{Introduction}\label{sec:introduction}

When stock markets are closed, investors continue to assess news about firm value but lack contemporaneous transaction prices for the underlying stocks (\citealp{r13,r0}). Equity perpetuals offer synthetic exposure to individual stocks and trade around the clock, generating observable prices throughout the weekend (\citealp{r4,r17,r8}). Whether these prices provide useful information about stock prices at reopening is central to assessing the informational role of 24/7 derivatives.

Weekend perpetual prices may reflect value-relevant information as well as trading demand and liquidity conditions. Once stock trading resumes, new transactions and quotes provide more timely price information about the underlying stocks. Evaluating weekend perpetual prices therefore requires distinguishing their ability to explain subsequent stock returns from the information they add after stock prices have been updated. We examine both as stock trading resumes in stages and identify when the associated stock price adjustments occur.

Recent studies examine the information in tokenized stocks and perpetuals on traditional assets during market closures (\citealp{r4,r9,r10}). We extend this evidence to a broader sample of U.S.-listed stocks and trading venues. We combine transaction data on equity perpetuals covering 421 stocks across 22 venues from February to November 2026 with high-frequency trades and quotes for the underlying stocks. The successive reopening of stock index futures, overnight stock trading, premarket trading, and the regular session allows us to trace stock price adjustments. Using the cumulative stock return at the regular-session open as a common dependent variable, we assess the incremental explanatory power of weekend perpetual returns after conditioning on cumulative stock returns at each stage.

Weekend perpetual prices contain information about subsequent stock returns before stock trading resumes. In the baseline sample, a one-percentage-point higher weekend perpetual return is associated with a 59.2-basis-point higher cumulative stock return at the regular-session open, on average. Perpetual returns measured before S\&P 500 index futures reopen are also significantly positively related to cumulative opening returns. The positive relationship for the full weekend return remains significant after controlling for market exposure based on stock index futures returns. In the expanded stock sample, the perpetual return coefficient remains positive when each trading venue is excluded in turn.

Stock price adjustments associated with weekend perpetual returns are already apparent early in the overnight session. In the common overnight and daytime sample, conditioning on cumulative stock returns five minutes after the overnight open reduces the incremental within-$R^2$ of weekend perpetual returns for cumulative opening returns from 7.12 to 0.07 percentage points. In the sample with stock prices available five minutes before the regular-session open, the remaining opening effect, conditional on the cumulative stock return at that time, passes an equivalence test with bounds of $\pm10$ basis points per percentage point of perpetual return.

Trade- and quote-level evidence further shows that early stock price adjustment continues after the first transaction. Weekend perpetual returns are significantly positively associated with stock price changes within the first minute with valid trades. Conditional on the weekend perpetual return and the cumulative stock return at the first transaction price, the within-minute stock return contains incremental information about the cumulative opening return.

\par\medskip\noindent\textbf{Related literature.}\quad This paper relates to three strands of the literature. The first concerns multimarket price discovery and differences in trading hours. \citet{r6} uses innovations in a common efficient price to measure the relative contributions of different markets to price discovery. Studies of stock and option markets show that option trading and prices contain information relevant to underlying stock values (\citealp{r5,r24}). Research on stock index futures and cash markets examines the information content of derivative prices and their lead--lag relations with cash prices (\citealp{r3,r25}).

Differences in trading hours also shape the roles of related markets in price discovery. \citet{r26} incorporate nonoverlapping trading hours into measures of cross-market price discovery. \citet{r27} find that trading in home stock markets improves the efficiency of opening prices for American depositary receipts. In cryptocurrency markets, \citet{r22} examines information generated during continuous spot trading and price adjustments when futures trading resumes. We study nonoverlapping trading in equity perpetuals and their underlying stocks, examining the relation between weekend price information and subsequent stock price adjustments.

Second, we contribute to the literature on trading sessions and opening price formation. Early research examines how information arrival, trading behavior, and trading mechanisms affect price formation across trading sessions (\citealp{r13,r0,r1}). Recent studies focus on the information content of overnight and premarket trading in U.S. stocks and the persistence of price changes in these sessions. \citet{r28} find that overnight trading is driven primarily by retail investors and examine the persistence of overnight price changes in the subsequent regular session. \citet{r29} finds that cumulative premarket returns gain explanatory power for eventual opening returns as premarket trading progresses and that some early price changes subsequently reverse.\footnote{From a market design perspective, \citet{r38} examine the role of market closures in coordinating liquidity and the economic consequences of extending trading hours.}

Research on opening price formation documents the informational value of preopening quotes, indicative prices, and opening auctions (\citealp{r30,r2,r11}). The market microstructure literature examines value-relevant information, transitory price components, and price pressure in transaction prices (\citealp{r12,r31,r32,r7}). Related work also documents persistence in overnight and intraday returns and reversals across trading sessions (\citealp{r33}). We extend this research to nonoverlapping trading in 24/7 equity perpetuals and their underlying stocks. We examine the timing of price adjustments as stock trading resumes in stages and use trade- and quote-level data to analyze price updates early in the reopening process.

Finally, our paper relates directly to the emerging literature on stock tokenization and 24/7 perpetuals. Research on cryptocurrency markets documents cross-venue arbitrage frictions and the role of perpetual contracts in price discovery (\citealp{r36,r35}).\footnote{The theoretical design of perpetual futures dates back to \citet{r23}. \citet{r34} further examine no-arbitrage pricing and funding rate mechanisms for modern perpetual contracts.} Recent studies of tokenized stocks examine their information content outside regular trading hours, trading activity, and price tracking errors (\citealp{r4,r17,r18,r19,r20}). \citet{r21} finds that traditional stock markets dominate long-run price discovery between tokenized and underlying stocks in the sample studied. \citet{r8} compares the execution mechanisms of equity-backed tokens and synthetic perpetual contracts and their relation to order flow in the underlying stock market.

As perpetual contracts expand to traditional financial assets, research has begun to examine price tracking and information content across asset classes (\citealp{r37}). Among the studies most closely related to ours, \citet{r9} examine the formation of perpetual reference prices when external reference prices cease updating, and \citet{r10} studies the predictive information in perpetuals on traditional assets during market closures and its incorporation at the open. Building on this work, we extend the evidence to a broader sample of U.S.-listed stocks and trading venues. Combining price records from overnight, premarket, and regular trading, we examine the information content of weekend perpetual prices, the timing of the associated stock price adjustments, and the incremental value of perpetual prices after stock prices have been updated.

The remainder of the paper is organized as follows. \secref{sec:data_methods} describes the data and methodology, \secref{sec:results} presents the main empirical results, \secref{sec:microstructure} examines microstructure evidence and heterogeneity, \secref{sec:robustness} investigates common market information and presents robustness tests, and \secref{sec:conclusion} concludes.

\section{Data and Methodology}\label{sec:data_methods}

\subsection{Institutional Setting and Reopening Schedule}

Equity perpetual contracts are derivatives on individual stock prices that typically have no fixed expiration date and provide synthetic equity exposure through margin trading. They generally support round-the-clock (24/7) trading, allowing transaction prices to form while the underlying stock market is closed.

\begin{figure}[H]
\centering
\includegraphics[width=.94\linewidth]{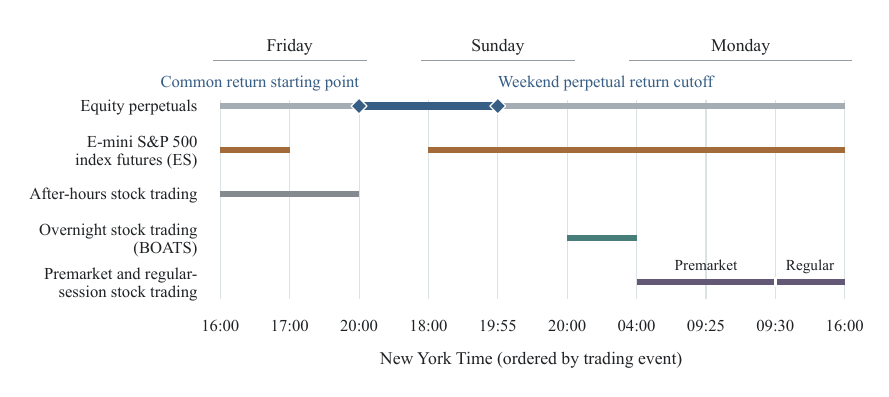}
\caption{Trading schedules for equity perpetuals, U.S. stocks, and stock index futures}\label{fig:1}
\begin{exhibitnotes}
The figure shows the trading schedules for weekends followed by a Monday trading day. All times are New York local time, and the horizontal axis follows the sequence of trading events. Diamonds mark the common return starting point (Friday 20:00) and the cutoff for weekend perpetual returns (Sunday 19:55).
\end{exhibitnotes}
\end{figure}

All times are New York local time unless otherwise stated. U.S. stocks and related derivatives resume trading in succession after the weekend. On weekends followed by a Monday trading session (hereafter, regular weekends), E-mini S\&P 500 futures (ES) reopen at 18:00 on Sunday, and Blue Ocean ATS (BOATS) opens for overnight stock trading at 20:00. Premarket trading and the regular trading session begin at 04:00 and 09:30 the following day, respectively. We refer to the start of the regular trading session as the regular-session open.\footnote{See the BOATS trading hours and CME contract specifications.} \figref{fig:1} illustrates this sequence.

The baseline analysis measures both perpetual and stock returns from 20:00 on Friday. Weekend perpetual returns end at 19:55 on Sunday, five minutes before the overnight open on regular weekends. We also use 18:00 on Sunday to distinguish perpetual returns before and after ES reopens.\footnote{When Monday is a market holiday, perpetual returns still end at 19:55 on Sunday. Overnight stock prices and prices at the regular-session open are determined using the actual reopening date and the preceding evening.}

\subsection{Data Sources and Sample Construction}

The sample covers weekend reopening events from February 2 through November 23, 2026. We match the corresponding equity perpetual trade records to U.S.-listed stock prices. Each observation pairs a stock with a reopening event and is referred to as a stock-week. The stocks comprise U.S.-listed common stocks and eligible American depositary receipts (ADRs). Perpetual trade data come primarily from the venues' official historical data. Daily and minute-level stock prices, individual trades, and quotes come from Alpaca. \appref{app:A2} describes the other market data and their sources.

We identify equity perpetuals on U.S.-listed securities using contract directories, listing announcements, and historical trade records.\footnote{Different listed share classes of the same issuer are treated as separate securities. Trading channels that share the same underlying market are treated as one venue. The contract directory is constructed from information available as of November 19, 2026, and includes listed contracts and delisted contracts with verifiable historical records.} Matching valid perpetual returns to the corresponding stock prices yields an expanded sample of 7,489 stock-weeks covering 421 stocks and 22 venues. In this sample, stock returns start at 16:00 on Friday, while perpetual returns still start at 20:00. To align the starting points of the two returns, the baseline analysis requires stock and perpetual prices at Friday 20:00 and complete historical stock characteristics. This sample contains 4,060 stock-weeks for 319 stocks. \appref{app:A1} describes the sample construction.

\subsection{Variable Definitions and Return Measurement}

We measure weekend perpetual returns and stock returns at successive stages after stock trading resumes. Let $i$ index stocks, $w$ reopening events, and $v$ trading venues. In the baseline analysis, $t_0$ and $t_1$ denote Friday 20:00 and Sunday 19:55, respectively, and $o$ denotes the regular-session open. Let $P^P_{ivw}(t)$ be the transaction price of the perpetual on stock $i$ at venue $v$. The stock-level weekend perpetual return is the equal-weighted average of log returns across venues with valid prices:
\begin{gather}
 W_{ivw}(a,b)=100\ln\frac{P^P_{ivw}(b)}{P^P_{ivw}(a)},\label{eq:venue_return}\\
 W_{iw}=\frac{1}{|\mathcal V_{iw}|}\sum_{v\in\mathcal V_{iw}}W_{ivw}(t_0,t_1).\label{eq:perp_return}
\end{gather}
Here, $\mathcal V_{iw}$ is the set of venues with valid transaction prices at both $t_0$ and $t_1$.

Let $P^S_{iw}(t)$ denote the underlying stock price. The cumulative stock return is
\begin{equation}\label{eq:stock_return}
 R_{iw}(t)=100\ln\frac{P^S_{iw}(t)}{P^S_{iw}(t_0)}.
\end{equation}
The corresponding interval return is $R_{iw}(a,b)=R_{iw}(b)-R_{iw}(a)$. We refer to $R_{iw}(o)$, the cumulative stock return at the regular-session open, as the cumulative opening return. All returns are expressed in percent.

The main stock price observation times are five minutes after the overnight open, five minutes after premarket trading begins, and five minutes before the regular-session open (09:25). The first two measure early price responses in the respective trading sessions. We measure the first overnight transaction price using the opening price of the first minute with valid trades within the first hour after the overnight open.

We use closing prices from five-minute perpetual bars and one-minute stock bars, except for the first overnight transaction price and the regular-session opening price. At each observation time, we select the most recent valid bar ending at or before that time. The bar must start no more than 15 minutes earlier for perpetuals and five minutes earlier for stocks.\footnote{The price observation lag is the time elapsed between the start of the bar and the observation time. \appref{app:A5} provides the price measurement details.} The regular-session opening price is the unadjusted daily opening price.

To examine when weekend price information emerges, we divide the weekend into four intervals at Saturday 20:00, Sunday 12:00, and Sunday 18:00. Cumulative and interval perpetual returns use a common set of venues with valid prices at all designated observation times. For the analysis of information before ES reopens, we also use venues selected from current and historical trade records up to and including Sunday 18:00. \apptabref{tab:A3} provides the selection rules.

The historical stock characteristics $H_{iw}$ comprise the Friday regular-session return, trailing 5-day and 20-day returns, trailing 20-day volatility, and historical mean log dollar volume. All are constructed using data through Friday, with dollar volume measured in U.S. dollars. \apptabref{tab:A3}, \apptabref{tab:A4}, and \appref{app:A5} provide the variable definitions and construction details.

\subsection{Descriptive Statistics}

\tabref{tab:1} reports descriptive statistics and correlations for the main variables. Using valid baseline-sample observations, the standard deviations of weekend perpetual returns and cumulative opening returns are 1.273\% and 3.069\%, respectively. The pairwise correlations of weekend perpetual returns with cumulative stock returns five minutes after the overnight open and at the regular-session open are 0.673 and 0.385. The correlation between cumulative stock returns five minutes before and at the regular-session open is 0.976.

\begin{table}[H]
\centering\setlength{\parskip}{0pt}
\caption{Descriptive statistics and correlations}\label{tab:1}
\begingroup\linespread{1.0}\fontsize{9}{10.5}\selectfont\setlength{\tabcolsep}{2.0pt}\renewcommand{\arraystretch}{1.0}
\begin{tabular}{@{}L{0.38\linewidth}*{6}{C{\dimexpr(\linewidth-0.38\linewidth-12\tabcolsep)/6\relax}}@{}}\toprule
\multicolumn{7}{@{}p{\linewidth}@{}}{\bfseries Panel A. Descriptive statistics}\\[2pt]
Variable & \grouphead{Stock-\\weeks} & Mean & Std. dev. & 25th pct. & Median & 75th pct.\\\midrule
\addlinespace[3pt]\multicolumn{7}{@{}p{\linewidth}@{}}{\bfseries Perpetual and stock returns (\%)}\\
\rowlabel{Weekend perpetual return} & 4,060 & \descnum{0}{320} & \descnum{1}{273} & \descnum{-0}{194} & \descnum{0}{164} & \descnum{0}{763}\\
\rowlabel{Pre-ES perpetual return} & 3,838 & \descnum{0}{229} & \descnum{1}{203} & \descnum{-0}{244} & \descnum{0}{114} & \descnum{0}{637}\\
\rowlabel{First overnight trade cum. return} & 3,603 & \descnum{0}{374} & \descnum{1}{339} & \descnum{-0}{138} & \descnum{0}{230} & \descnum{0}{900}\\
\rowlabel{Overnight +5 min cum. return} & 2,859 & \descnum{0}{385} & \descnum{1}{672} & \descnum{-0}{258} & \descnum{0}{300} & \descnum{1}{069}\\
\rowlabel{09:25 cum. return} & 3,708 & \descnum{0}{568} & \descnum{3}{139} & \descnum{-0}{912} & \descnum{0}{212} & \descnum{1}{637}\\
\rowlabel{Cumulative opening return} & 4,060 & \descnum{0}{481} & \descnum{3}{069} & \descnum{-0}{910} & \descnum{0}{165} & \descnum{1}{489}\\
\rowlabel{09:25 to regular-session open return} & 3,708 & \descnum{-0}{036} & \descnum{0}{687} & \descnum{-0}{378} & \descnum{-0}{015} & \descnum{0}{331}\\
\rowlabel{Regular-session open-to-close return} & 4,060 & \descnum{-0}{014} & \descnum{3}{876} & \descnum{-1}{858} & \descnum{-0}{074} & \descnum{1}{693}\\
\addlinespace[3pt]\multicolumn{7}{@{}p{\linewidth}@{}}{\bfseries Historical stock characteristics}\\
\rowlabel{Friday regular-session return (\%)} & 4,060 & \descnum{-0}{707} & \descnum{3}{959} & \descnum{-2}{631} & \descnum{-0}{495} & \descnum{1}{194}\\
\rowlabel{Trailing 5-day return (\%)} & 4,060 & \descnum{-1}{219} & \descnum{11}{296} & \descnum{-6}{316} & \descnum{-0}{769} & \descnum{4}{090}\\
\rowlabel{Trailing 20-day return (\%)} & 4,060 & \descnum{-2}{286} & \descnum{22}{067} & \descnum{-13}{685} & \descnum{-1}{980} & \descnum{9}{026}\\
\rowlabel{Trailing 20-day volatility (\%)} & 4,060 & \descnum{4}{275} & \descnum{2}{608} & \descnum{2}{420} & \descnum{3}{837} & \descnum{5}{593}\\
\rowlabel{Historical mean log dollar volume} & 4,060 & \descnum{20}{845} & \descnum{1}{543} & \descnum{20}{018} & \descnum{20}{931} & \descnum{21}{730}\\
\addlinespace[3pt]\multicolumn{7}{@{}p{\linewidth}@{}}{\bfseries Trading characteristics and market exposure}\\
\rowlabel{Valid-return venues} & 4,060 & \descnum{4}{729} & \descnum{3}{179} & \descnum{2}{000} & \descnum{4}{000} & \descnum{7}{000}\\
\rowlabel{Cross-venue return dispersion (\%)} & 3,195 & \descnum{0}{191} & \descnum{0}{223} & \descnum{0}{073} & \descnum{0}{125} & \descnum{0}{220}\\
\rowlabel{ES market exposure (\%)} & 3,091 & \descnum{0}{295} & \descnum{1}{545} & \descnum{-0}{331} & \descnum{0}{128} & \descnum{0}{573}\\
\midrule\end{tabular}\endgroup
\par\nointerlineskip
\begingroup\linespread{1.0}\fontsize{9}{10.5}\selectfont\setlength{\tabcolsep}{1.6pt}\renewcommand{\arraystretch}{1.0}
\begin{tabular}{@{}L{0.34\linewidth}*{8}{C{\dimexpr(\linewidth-0.34\linewidth-16\tabcolsep)/8\relax}}@{}}
\multicolumn{9}{@{}p{\linewidth}@{}}{\bfseries Panel B. Correlation matrix}\\[2pt]
Variable & 1 & 2 & 3 & 4 & 5 & 6 & 7 & 8\\\midrule
\rowlabel{1 Weekend perpetual return} & \descnum{1}{000} &  &  &  &  &  &  & \\
\rowlabel{2 Pre-ES perpetual return} & \descnum{0}{833} & \descnum{1}{000} &  &  &  &  &  & \\
\rowlabel{3 First overnight trade cum. return} & \descnum{0}{627} & \descnum{0}{473} & \descnum{1}{000} &  &  &  &  & \\
\rowlabel{4 Overnight +5 min cum. return} & \descnum{0}{673} & \descnum{0}{527} & \descnum{0}{802} & \descnum{1}{000} &  &  &  & \\
\rowlabel{5 09:25 cum. return} & \descnum{0}{421} & \descnum{0}{235} & \descnum{0}{482} & \descnum{0}{533} & \descnum{1}{000} &  &  & \\
\rowlabel{6 Cumulative opening return} & \descnum{0}{385} & \descnum{0}{210} & \descnum{0}{453} & \descnum{0}{513} & \descnum{0}{976} & \descnum{1}{000} &  & \\
\rowlabel{7 Regular-session open-to-close return} & \descnum{-0}{011} & \descnum{0}{045} & \descnum{-0}{056} & \descnum{-0}{012} & \descnum{-0}{059} & \descnum{-0}{055} & \descnum{1}{000} & \\
\rowlabel{8 Cross-venue return dispersion} & \descnum{-0}{090} & \descnum{-0}{043} & \descnum{-0}{060} & \descnum{-0}{098} & \descnum{-0}{088} & \descnum{-0}{098} & \descnum{0}{066} & \descnum{1}{000}\\
\bottomrule\end{tabular}\endgroup
\begin{exhibitnotes}
Panel A reports descriptive statistics using the valid observations for each variable in the baseline sample. The cumulative return at the first overnight trade is measured using the opening price of the first minute with valid trades. Returns, daily return volatility, cross-venue return dispersion, and ES market exposure are expressed in \%. Panel B reports Pearson correlations based on pairwise valid observations.
\end{exhibitnotes}
\end{table}

\subsection{Empirical Specifications and Statistical Inference}

\paragraph{Baseline regression and dynamic price responses.}
We relate weekend perpetual returns to cumulative stock returns after trading resumes using the following model:
\begin{equation}\label{eq:baseline}
 R_{iw}(t)=\alpha_i+\delta_w+\beta_t W_{iw}
 +\gamma_t'H_{iw}+\varepsilon_{iwt}.
\end{equation}
Here, $\alpha_i$ and $\delta_w$ are stock and actual reopening-date fixed effects, respectively. The coefficient $\beta_t$ measures the relation between weekend perpetual returns and cumulative stock returns at time $t$, conditional on historical stock characteristics $H_{iw}$. The baseline dependent variable is the cumulative opening return. We also estimate regressions for cumulative and interval returns during overnight, premarket, and regular trading. We use common stock-week samples to compare price responses across stages and all valid observations at each time to assess the effect of the sample restrictions.

\paragraph{Timing of weekend price information.}
To distinguish information arising at different stages of the weekend, we replace the full weekend return with returns over the four intervals:
\begin{equation}\label{eq:segments}
 R_{iw}(o)=\alpha_i+\delta_w+\sum_{k=1}^{4}b_k W^{(k)}_{iw}
 +\gamma'H_{iw}+\epsilon_{iw}.
\end{equation}
The variable $W^{(k)}_{iw}$ is the perpetual return in interval $k$. Its coefficient $b_k$ measures the relation with the cumulative opening return, conditional on perpetual returns in the other intervals. We test differences across the coefficients. We also combine the first three interval returns and include the resulting return jointly with the Sunday 18:00--19:55 return to compare information before and after ES reopens.

\paragraph{Incremental information in perpetual returns.}
The incremental information model measures the explanatory power of perpetual returns beyond stock price changes already observed. We keep the cumulative opening return as the dependent variable and include the cumulative stock return at each observation time in a separate regression:
\begin{equation}\label{eq:remaining}
 R_{iw}(o)=\alpha_i+\delta_w+c_t W_{iw}
 +\theta_t R_{iw}(t)+\gamma_t'H_{iw}+\nu_{iwt}.
\end{equation}
The coefficient $c_t$ measures the relation between weekend perpetual returns and cumulative opening returns conditional on the cumulative stock return at time $t$. A one-percentage-point higher weekend perpetual return is associated with a conditional difference of $100c_t$ basis points in the cumulative opening return. We call this the remaining opening effect.

We measure the incremental explanatory power of perpetual returns as the difference in within-$R^2$ between models with and without $W_{iw}$, holding the dependent variable, sample, fixed effects, and other regressors constant. Incremental within-$R^2$ is reported in percentage points; the \% symbol used for this measure in tables and figures denotes percentage points.

We further assess the economic magnitude of the remaining opening effect using two one-sided tests (TOST; \citealp{r16}). The equivalence bounds are $\pm5$, $\pm10$, and $\pm20$ basis points per one-percentage-point higher weekend perpetual return. With the same standard errors and critical values, equivalence at the 5\% significance level requires the corresponding 90\% confidence interval to lie entirely within the specified bounds.\footnote{We use 5, 10, and 20 basis points as benchmarks for economic magnitude to assess how the equivalence conclusions depend on the width of the bounds. These benchmarks do not represent transaction cost thresholds.}

\FloatBarrier
\section{Empirical Results}\label{sec:results}

\subsection{Information Content of Weekend Perpetual Prices}

We first estimate the association between weekend perpetual returns and cumulative opening returns. \tabref{tab:2} reports estimates of \eqnref{eq:baseline}. The coefficient on weekend perpetual returns ranges from 0.589 to 0.594 as historical stock characteristics are added, and is significant at the 1\% level in every specification. With the full set of controls, the coefficient is 0.592: a one-percentage-point higher weekend perpetual return is associated with a 59.2-basis-point higher cumulative opening return, on average.

Using the descriptive statistics in \tabref{tab:1}, a one-standard-deviation difference in weekend perpetual returns corresponds to a difference of approximately 75.4 basis points in cumulative opening returns, or 24.6\% of their standard deviation. Adding weekend perpetual returns to a model with historical stock characteristics and fixed effects raises within-$R^2$ by 6.994 percentage points. Together, these results indicate that weekend perpetual prices contain information about cumulative opening returns before stock trading resumes.

\begin{table}[H]
\centering\setlength{\parskip}{0pt}
\caption{Weekend perpetual returns and cumulative opening returns}\label{tab:2}
\begingroup
\linespread{1.0}\fontsize{9.5}{11}\selectfont
\setlength{\tabcolsep}{2.0pt}
\renewcommand{\arraystretch}{1.0}
\begin{tabular}{@{}L{0.37\linewidth}*{6}{C{\dimexpr(\linewidth-0.37\linewidth-12\tabcolsep)/6\relax}}@{}}
\toprule
 & \multicolumn{6}{C{\dimexpr ((\linewidth-0.37\linewidth-12\tabcolsep)/6)*6+10\tabcolsep\relax}@{}}{Dependent variable: Cumulative opening return}\\
\cmidrule(lr){2-7}
Variable & (1) & (2) & (3) & (4) & (5) & (6)\\
\midrule
\rowlabel{Weekend perpetual return} & \regpair{\regcoef{0}{592}{***}}{\regse{0}{179}} & \regpair{\regcoef{0}{589}{***}}{\regse{0}{174}} & \regpair{\regcoef{0}{593}{***}}{\regse{0}{170}} & \regpair{\regcoef{0}{594}{***}}{\regse{0}{170}} & \regpair{\regcoef{0}{590}{***}}{\regse{0}{166}} & \regpair{\regcoef{0}{592}{***}}{\regse{0}{166}}\\
\rowlabel{Friday regular-session return} &  & \regpair{\regcoef{-0}{021}{}}{\regse{0}{048}} & \regpair{\regcoef{0}{003}{}}{\regse{0}{042}} & \regpair{\regcoef{0}{002}{}}{\regse{0}{041}} & \regpair{\regcoef{0}{004}{}}{\regse{0}{039}} & \regpair{\regcoef{0}{005}{}}{\regse{0}{039}}\\
\rowlabel{Trailing 5-day return} &  &  & \regpair{\regcoef{-0}{021}{*}}{\regse{0}{013}} & \regpair{\regcoef{-0}{025}{*}}{\regse{0}{014}} & \regpair{\regcoef{-0}{025}{*}}{\regse{0}{014}} & \regpair{\regcoef{-0}{026}{*}}{\regse{0}{014}}\\
\rowlabel{Trailing 20-day return} &  &  &  & \regpair{\regcoef{0}{004}{}}{\regse{0}{007}} & \regpair{\regcoef{0}{002}{}}{\regse{0}{008}} & \regpair{\regcoef{0}{003}{}}{\regse{0}{008}}\\
\rowlabel{Trailing 20-day volatility} &  &  &  &  & \regpair{\regcoef{0}{049}{}}{\regse{0}{072}} & \regpair{\regcoef{0}{044}{}}{\regse{0}{072}}\\
\rowlabel{Historical mean log dollar volume} & NO & NO & NO & NO & NO & YES\\
\rowlabel{Stock fixed effects} & YES & YES & YES & YES & YES & YES\\
\rowlabel{Reopening-date fixed effects} & YES & YES & YES & YES & YES & YES\\

\addlinespace[2pt]
\rowlabel{Stock-weeks} & 4,060 & 4,060 & 4,060 & 4,060 & 4,060 & 4,060\\
\rowlabel{Stocks} & 319 & 319 & 319 & 319 & 319 & 319\\
\rowlabel{Within-\ensuremath{R^2}} & \regstat{0}{070} & \regstat{0}{071} & \regstat{0}{078} & \regstat{0}{079} & \regstat{0}{080} & \regstat{0}{080}\\
\rowlabel{Perpetual incr. within-\ensuremath{R^2} (\%)} & \regstat{7}{044} & \regstat{6}{947} & \regstat{7}{055} & \regstat{7}{072} & \regstat{6}{970} & \regstat{6}{994}\\
\bottomrule\end{tabular}
\endgroup

\begin{exhibitnotes}
The dependent variable is the cumulative opening return (\%).
All columns use the same stock-week sample, include stock and reopening-date fixed effects, and add historical stock characteristics sequentially.
Column (6) also controls for historical mean log dollar volume; its coefficient is not reported.
\incrnote{}\regnote{}
\end{exhibitnotes}
\end{table}

\subsection{Information in Perpetual Prices before Stock Index Futures Reopen}

To assess the information in perpetual prices before stock index futures reopen, we estimate separate regressions using perpetual returns ending at 18:00 and 19:55 on Sunday. Panel A of \tabref{tab:3} relates each measure to subsequent cumulative stock returns, holding the stock-week and venue samples fixed within each column. Returns measured before ES reopens are significantly positively associated with subsequent cumulative stock returns during overnight and premarket trading and at the regular-session open. The coefficient in the cumulative opening return regression is 0.361, significant at the 5\% level. Extending the perpetual return window to 19:55 raises the corresponding coefficient to 0.627 and incremental within-$R^2$ from 2.441 to 7.919 percentage points.

We also select venues using trade records available by 18:00 on Sunday, aligning venue selection with the perpetual return cutoff. In Panel B of \tabref{tab:3}, the cumulative opening return coefficients are 0.246 for venues with valid returns through 18:00 and 0.288 for historically active venues, both significant at the 5\% level. In a common sample of 3,436 stock-weeks, the coefficients range from 0.313 to 0.318 across the three venue selection rules and are all significant at the 5\% level (\apptabref{tab:B1}, Panel A). The consistent results across venue selection rules indicate that perpetual prices formed before ES reopens already contain information about subsequent stock returns.

\begin{table}[H]
\centering\setlength{\parskip}{0pt}
\caption{Information in perpetual prices before stock index futures reopen}\label{tab:3}

\begingroup
\linespread{1.0}\fontsize{9.5}{11}\selectfont
\setlength{\tabcolsep}{2.2pt}
\renewcommand{\arraystretch}{1.0}

\begin{tabular}{@{}L{0.38\linewidth}*{4}{C{\dimexpr(\linewidth-0.38\linewidth-8\tabcolsep)/4\relax}}@{}}
\toprule
\multicolumn{5}{@{}p{\linewidth}@{}}{\bfseries Panel A. Perpetual returns at different cutoffs}\\[2pt]
Variable & \grouphead{First overnight\\price} & \grouphead{Overnight\\+5 min} & 09:25 & \grouphead{Regular-session\\open}\\
\midrule
\rowlabel{Pre-ES perpetual return} & \regpair{\regcoef{0}{422}{***}}{\regse{0}{066}} & \regpair{\regcoef{0}{547}{***}}{\regse{0}{064}} & \regpair{\regcoef{0}{400}{***}}{\regse{0}{146}} & \regpair{\regcoef{0}{361}{**}}{\regse{0}{142}}\\
\rowlabel{Weekend perpetual return} & \regpair{\regcoef{0}{498}{***}}{\regse{0}{058}} & \regpair{\regcoef{0}{641}{***}}{\regse{0}{053}} & \regpair{\regcoef{0}{681}{***}}{\regse{0}{161}} & \regpair{\regcoef{0}{627}{***}}{\regse{0}{169}}\\
\rowlabel{18:00 incr. within-\ensuremath{R^2} (\%)} & \regstat{16}{623} & \regstat{22}{647} & \regstat{3}{079} & \regstat{2}{441}\\
\rowlabel{19:55 incr. within-\ensuremath{R^2} (\%)} & \regstat{24}{901} & \regstat{33}{031} & \regstat{9}{684} & \regstat{7}{919}\\

\addlinespace[2pt]
\rowlabel{Stock-weeks} & 3,412 & 2,704 & 3,511 & 3,838\\
\rowlabel{Stocks} & 261 & 215 & 275 & 314\\

\midrule\end{tabular}
\endgroup

\par\nointerlineskip

\begingroup
\linespread{1.0}\fontsize{9.5}{11}\selectfont
\setlength{\tabcolsep}{2.2pt}
\renewcommand{\arraystretch}{1.0}

\begin{tabular}{@{}L{0.38\linewidth}*{4}{C{\dimexpr(\linewidth-0.38\linewidth-8\tabcolsep)/4\relax}}@{}}
\multicolumn{5}{@{}p{\linewidth}@{}}{\bfseries Panel B. Perpetual returns using venues selected before ES reopens}\\[2pt]
Variable & \grouphead{First overnight\\price} & \grouphead{Overnight\\+5 min} & 09:25 & \grouphead{Regular-session\\open}\\
\midrule

\addlinespace[3pt]
\multicolumn{5}{@{}p{\linewidth}@{}}{\bfseries Venues with valid returns through 18:00}\\
\rowlabel{Pre-ES perpetual return} & \regpair{\regcoef{0}{398}{***}}{\regse{0}{065}} & \regpair{\regcoef{0}{526}{***}}{\regse{0}{064}} & \regpair{\regcoef{0}{282}{**}}{\regse{0}{116}} & \regpair{\regcoef{0}{246}{**}}{\regse{0}{111}}\\

\addlinespace[2pt]
\rowlabel{Stock-weeks} & 3,439 & 2,734 & 3,539 & 3,881\\
\rowlabel{Stocks} & 265 & 220 & 281 & 319\\
\rowlabel{Within-\ensuremath{R^2}} & \regstat{0}{168} & \regstat{0}{221} & \regstat{0}{027} & \regstat{0}{022}\\
\rowlabel{Perpetual incr. within-\ensuremath{R^2} (\%)} & \regstat{15}{449} & \regstat{21}{482} & \regstat{1}{917} & \regstat{1}{416}\\

\addlinespace[3pt]
\multicolumn{5}{@{}p{\linewidth}@{}}{\bfseries Historically active venues}\\
\rowlabel{Pre-ES perpetual return} & \regpair{\regcoef{0}{429}{***}}{\regse{0}{066}} & \regpair{\regcoef{0}{565}{***}}{\regse{0}{063}} & \regpair{\regcoef{0}{328}{**}}{\regse{0}{126}} & \regpair{\regcoef{0}{288}{**}}{\regse{0}{122}}\\

\addlinespace[2pt]
\rowlabel{Stock-weeks} & 3,157 & 2,519 & 3,245 & 3,542\\
\rowlabel{Stocks} & 248 & 206 & 259 & 292\\
\rowlabel{Within-\ensuremath{R^2}} & \regstat{0}{192} & \regstat{0}{261} & \regstat{0}{033} & \regstat{0}{028}\\
\rowlabel{Perpetual incr. within-\ensuremath{R^2} (\%)} & \regstat{17}{955} & \regstat{25}{175} & \regstat{2}{571} & \regstat{1}{944}\\

\bottomrule\end{tabular}
\endgroup

\begin{exhibitnotes}
The dependent variable in each column is the cumulative stock return at the indicated time (\%). The first overnight price is the opening price of the first minute with valid overnight trades.
All models control for historical stock characteristics and include stock and reopening-date fixed effects.
Panel A enters the two perpetual return measures in separate regressions. Within each column, both regressions use the same stock-weeks and venues with valid returns over the full window.
The return cutoffs are Sunday 18:00 and 19:55, respectively, and incremental within-$R^2$ corresponds to perpetual returns at each cutoff.
Panel B selects venues using trade records through Sunday 18:00.
Historically active venues must also have valid prices at the specified start and end times in at least two of the preceding four calendar weekends.
The Panel B sample ends on November 16, 2026.
\incrnote{}\regnote{}
\end{exhibitnotes}
\end{table}

\subsection{Timing of Weekend Price Information}

The association between perpetual returns and subsequent cumulative opening returns may differ across weekend intervals. Panel A of \tabref{tab:4} reports a baseline regression that jointly includes four interval returns; all four coefficients are significantly positive. With the full set of historical stock characteristics, the coefficients for the first and final intervals are 0.480 and 1.132. Their difference, 0.652, is significant at the 1\% level. For a one-percentage-point difference in perpetual returns, the associated difference in cumulative opening returns is 65.2 basis points larger in the final interval than in the first. In the common sample with industry classifications, the coefficient difference is 0.578 with reopening-date fixed effects and 0.411 with industry-by-reopening-date fixed effects, both significant at the 5\% level (\apptabref{tab:B1}, Panel B).

Splitting perpetual returns at Sunday 18:00 yields coefficients of 0.508 before ES reopens and 1.121 afterward. Estimates for regular weekends are similar (\tabref{tab:4}, Panel B). Conditional on perpetual returns after ES reopens, returns formed before the reopening remain significantly positively associated with cumulative opening returns.

\begin{figure}[H]
\centering
\includegraphics[width=.94\linewidth]{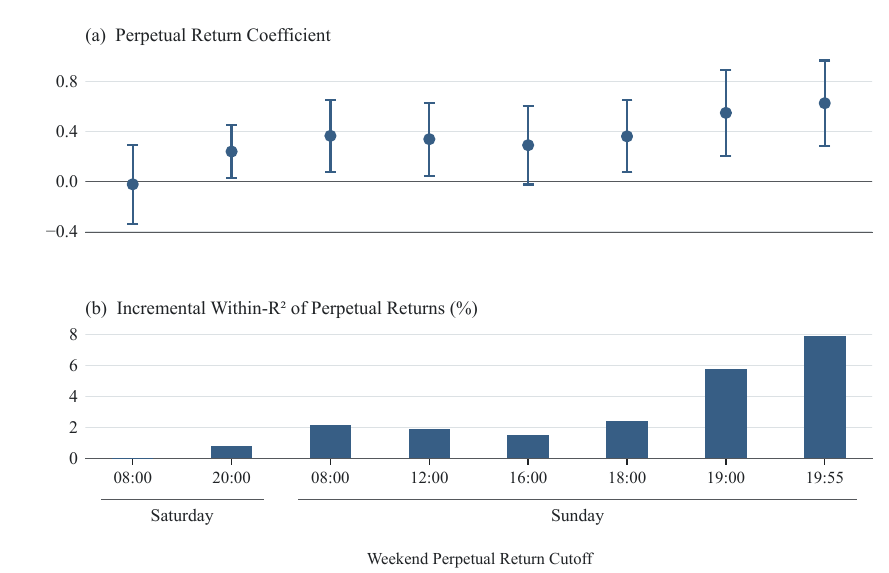}
\caption{Timing of information in weekend perpetual prices}\label{fig:2}

\begin{exhibitnotes}
The figure relates cumulative perpetual returns at different weekend cutoffs to cumulative opening returns, using the same 3,838 stock-weeks and trading venues. All models control for historical stock characteristics and include stock and reopening-date fixed effects. The upper panel reports perpetual return coefficients and 95\% confidence intervals based on two-way clustering; the lower panel reports the corresponding incremental within-$R^2$ (\%). \incrnote{} The horizontal axis is ordered by the perpetual return cutoff.
\end{exhibitnotes}
\end{figure}

\figref{fig:2} plots the coefficients and incremental explanatory power of cumulative perpetual returns ending at different times over the weekend. The coefficient is already significantly positive for returns ending at Saturday 20:00, and both the coefficients and incremental within-$R^2$ generally rise thereafter (\apptabref{tab:B1}, Panel C). Together with the four-interval regressions, these estimates show that prices formed early in the weekend are informative and that the conditional association with cumulative opening returns is stronger for the final interval than for the first. In a local comparison around the ES reopening, however, the coefficients for the two adjacent 115-minute intervals do not differ significantly (\apptabref{tab:B1}, Panel D).

\begin{table}[H]
\centering\setlength{\parskip}{0pt}
\caption{Perpetual returns across weekend intervals and cumulative opening returns}\label{tab:4}

\begingroup
\linespread{1.0}\fontsize{9}{10.5}\selectfont
\setlength{\tabcolsep}{2.2pt}
\renewcommand{\arraystretch}{1.0}

\begin{tabular}{@{}L{0.38\linewidth}*{4}{C{\dimexpr(\linewidth-0.38\linewidth-8\tabcolsep)/4\relax}}@{}}
\toprule
\multicolumn{5}{@{}p{\linewidth}@{}}{\bfseries Panel A. Perpetual returns across weekend intervals}\\[2pt]
Variable & (1) & (2) & (3) & (4)\\
\midrule
\rowlabel{Fri. 20:00 to Sat. 20:00} & \regpair{\regcoef{0}{466}{***}}{\regse{0}{120}} & \regpair{\regcoef{0}{477}{***}}{\regse{0}{119}} & \regpair{\regcoef{0}{480}{***}}{\regse{0}{120}} & \regpair{\regcoef{0}{335}{***}}{\regse{0}{093}}\\
\rowlabel{Sat. 20:00 to Sun. 12:00} & \regpair{\regcoef{0}{500}{**}}{\regse{0}{217}} & \regpair{\regcoef{0}{504}{**}}{\regse{0}{209}} & \regpair{\regcoef{0}{502}{**}}{\regse{0}{200}} & \regpair{\regcoef{0}{365}{**}}{\regse{0}{158}}\\
\rowlabel{Sun. 12:00 to 18:00} & \regpair{\regcoef{0}{601}{***}}{\regse{0}{222}} & \regpair{\regcoef{0}{607}{***}}{\regse{0}{220}} & \regpair{\regcoef{0}{606}{***}}{\regse{0}{216}} & \regpair{\regcoef{0}{385}{**}}{\regse{0}{181}}\\
\rowlabel{Sun. 18:00 to 19:55} & \regpair{\regcoef{1}{156}{***}}{\regse{0}{302}} & \regpair{\regcoef{1}{135}{***}}{\regse{0}{274}} & \regpair{\regcoef{1}{132}{***}}{\regse{0}{268}} & \regpair{\regcoef{0}{746}{***}}{\regse{0}{208}}\\
\rowlabel{Final minus first coefficient} & \regpair{\regcoef{0}{690}{**}}{\regse{0}{257}} & \regpair{\regcoef{0}{658}{***}}{\regse{0}{238}} & \regpair{\regcoef{0}{652}{***}}{\regse{0}{235}} & \regpair{\regcoef{0}{411}{**}}{\regse{0}{164}}\\
\rowlabel{Coefficient difference $p$-value} & \regstat{0}{0107} & \regstat{0}{0086} & \regstat{0}{0083} & \regstat{0}{0165}\\

\addlinespace[2pt]
\rowlabel{Historical returns} & NO & YES & YES & YES\\
\rowlabel{Historical volatility and dollar volume} & NO & NO & YES & YES\\
\rowlabel{Industry\ensuremath{\times}reopening-date FE} & NO & NO & NO & YES\\

\addlinespace[2pt]
\rowlabel{Stock-weeks} & 3,838 & 3,838 & 3,838 & 3,174\\
\rowlabel{Stocks} & 314 & 314 & 314 & 172\\
\rowlabel{Within-\ensuremath{R^2}} & \regstat{0}{104} & \regstat{0}{112} & \regstat{0}{113} & \regstat{0}{058}\\

\midrule\end{tabular}
\endgroup

\par\nointerlineskip

\begingroup
\linespread{1.0}\fontsize{9}{10.5}\selectfont
\setlength{\tabcolsep}{2.0pt}
\renewcommand{\arraystretch}{1.0}

\begin{tabular}{@{}L{0.38\linewidth}*{6}{C{\dimexpr(\linewidth-0.38\linewidth-12\tabcolsep)/6\relax}}@{}}
\multicolumn{7}{@{}p{\linewidth}@{}}{\bfseries Panel B. Perpetual returns before and after ES reopens}\\[2pt]
 & \multicolumn{3}{C{\dimexpr ((\linewidth-0.38\linewidth-12\tabcolsep)/6)*3+4\tabcolsep\relax}}{All weekends} & \multicolumn{3}{C{\dimexpr ((\linewidth-0.38\linewidth-12\tabcolsep)/6)*3+4\tabcolsep\relax}@{}}{Regular weekends}\\
\cmidrule(lr){2-4}
\cmidrule(lr){5-7}
Variable & (1) & (2) & (3) & (4) & (5) & (6)\\
\midrule
\rowlabel{Pre-ES perpetual return} & \regpair{\regcoef{0}{498}{***}}{\regse{0}{147}} & \regpair{\regcoef{0}{507}{***}}{\regse{0}{142}} & \regpair{\regcoef{0}{508}{***}}{\regse{0}{140}} & \regpair{\regcoef{0}{535}{***}}{\regse{0}{152}} & \regpair{\regcoef{0}{543}{***}}{\regse{0}{146}} & \regpair{\regcoef{0}{544}{***}}{\regse{0}{143}}\\
\rowlabel{Post-ES perpetual return} & \regpair{\regcoef{1}{145}{***}}{\regse{0}{301}} & \regpair{\regcoef{1}{124}{***}}{\regse{0}{273}} & \regpair{\regcoef{1}{121}{***}}{\regse{0}{267}} & \regpair{\regcoef{1}{212}{***}}{\regse{0}{302}} & \regpair{\regcoef{1}{191}{***}}{\regse{0}{271}} & \regpair{\regcoef{1}{186}{***}}{\regse{0}{264}}\\

\addlinespace[2pt]
\rowlabel{Historical returns} & NO & YES & YES & NO & YES & YES\\
\rowlabel{Historical volatility and dollar volume} & NO & NO & YES & NO & NO & YES\\

\addlinespace[2pt]
\rowlabel{Stock-weeks} & 3,838 & 3,838 & 3,838 & 3,615 & 3,615 & 3,615\\
\rowlabel{Stocks} & 314 & 314 & 314 & 314 & 314 & 314\\
\rowlabel{Within-\ensuremath{R^2}} & \regstat{0}{1034} & \regstat{0}{1116} & \regstat{0}{1127} & \regstat{0}{1162} & \regstat{0}{1249} & \regstat{0}{1263}\\

\bottomrule\end{tabular}
\endgroup

\begin{exhibitnotes}
The dependent variable is the cumulative opening return (\%).
Panel A jointly includes perpetual returns from four nonoverlapping weekend intervals and reports the difference between the final and first coefficients and a test of that difference.
Column (4) uses industry-by-reopening-date fixed effects based on two-digit Standard Industrial Classification (SIC) codes.
Panel B splits perpetual returns at Sunday 18:00. The regular-weekend sample excludes weekends followed by a Monday market holiday.
All models include stock fixed effects and the corresponding date fixed effects. Historical returns comprise the Friday regular-session return and trailing 5-day and 20-day returns; historical volatility and dollar volume are trailing 20-day volatility and historical mean log dollar volume.
\regnote{}
\end{exhibitnotes}
\end{table}

\subsection{Stock Price Responses after Reopening}

\begin{figure}[H]
\centering
\includegraphics[width=.94\linewidth]{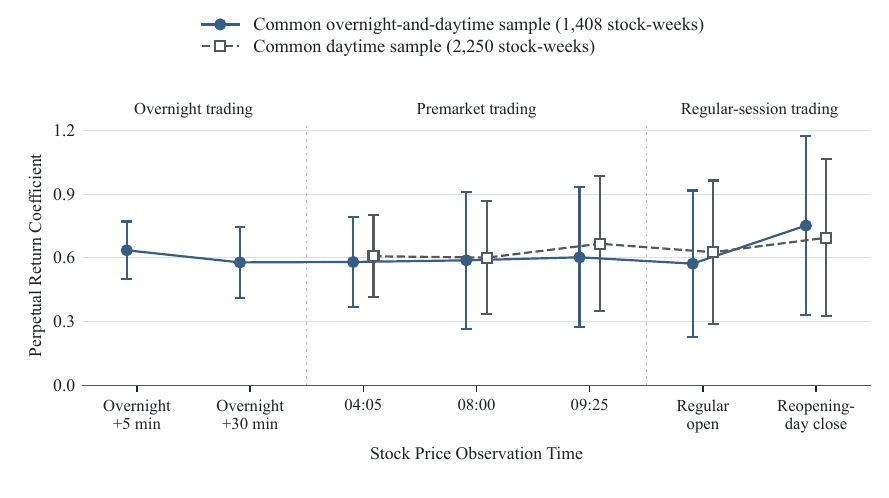}
\caption{Coefficients for cumulative stock returns after reopening}\label{fig:3}

\begin{exhibitnotes}
The figure reports perpetual return coefficients for the common overnight and daytime sample (1,408 stock-weeks) and the common daytime sample (2,250 stock-weeks). The dependent variable is the cumulative stock return at each indicated time. All models control for historical stock characteristics and include stock and reopening-date fixed effects. Stock-weeks are held fixed within each sample. Error bars show 95\% confidence intervals based on two-way clustering, and the horizontal axis is ordered by the stock price observation time.
\end{exhibitnotes}
\end{figure}

To trace stock price adjustment after reopening, we estimate stock return regressions for overnight, premarket, and regular trading. Panel A of \tabref{tab:5} and \figref{fig:3} show that, in the common overnight and daytime sample of 1,408 stock-weeks, the perpetual return coefficients are 0.636 for cumulative stock returns five minutes after the overnight open and 0.573 for cumulative opening returns. The coefficients at all reported times are significant at the 1\% level. Stock price adjustments associated with weekend perpetual returns are therefore already evident early in overnight trading.

Using all valid observations at each time yields a coefficient of 0.619 five minutes after the overnight open, close to the common-sample estimate. For cumulative stock returns five minutes after premarket trading begins, the coefficients are 0.608 in the common daytime sample and 0.602 using all valid observations (\tabref{tab:5}, Panels A and B). \apptabref{tab:B2} reports estimates for additional observation times.

In the common overnight and daytime sample, the coefficient on weekend perpetual returns is $-0.063$ in the regression of stock returns from five minutes after the overnight open to the regular-session open, and is not statistically significant. The coefficients for the other reported return intervals are also insignificant (\tabref{tab:5}, Panel C). By contrast, the positive association between weekend perpetual returns and cumulative stock returns remains significant at the regular-session open and the close of the reopening day.

\begin{table}[H]
\centering\setlength{\parskip}{0pt}
\caption{Cumulative and interval stock returns after reopening}\label{tab:5}

\begingroup
\linespread{1.0}\fontsize{9.5}{11}\selectfont
\setlength{\tabcolsep}{2.0pt}
\renewcommand{\arraystretch}{1.0}

\begin{tabular}{@{}L{0.38\linewidth}*{6}{C{\dimexpr(\linewidth-0.38\linewidth-12\tabcolsep)/6\relax}}@{}}
\toprule
\multicolumn{7}{@{}p{\linewidth}@{}}{\bfseries Panel A. Cumulative return regressions in common samples}\\[2pt]
 & \multicolumn{2}{C{\dimexpr ((\linewidth-0.38\linewidth-12\tabcolsep)/6)*2+2\tabcolsep\relax}}{Overnight trading} & \multicolumn{2}{C{\dimexpr ((\linewidth-0.38\linewidth-12\tabcolsep)/6)*2+2\tabcolsep\relax}}{Premarket trading} & \multicolumn{2}{C{\dimexpr ((\linewidth-0.38\linewidth-12\tabcolsep)/6)*2+2\tabcolsep\relax}@{}}{Regular trading}\\
\cmidrule(lr){2-3}
\cmidrule(lr){4-5}
\cmidrule(lr){6-7}
Sample & \grouphead{Overnight\\+5 min} & \grouphead{Overnight\\+30 min} & 04:05 & 09:25 & \grouphead{Regular\\open} & \grouphead{Reopening\\day close}\\
\midrule
\rowlabel{Common overnight and daytime sample} & \regpair{\regcoef{0}{636}{***}}{\regse{0}{067}} & \regpair{\regcoef{0}{579}{***}}{\regse{0}{083}} & \regpair{\regcoef{0}{581}{***}}{\regse{0}{105}} & \regpair{\regcoef{0}{603}{***}}{\regse{0}{163}} & \regpair{\regcoef{0}{573}{***}}{\regse{0}{170}} & \regpair{\regcoef{0}{752}{***}}{\regse{0}{208}}\\
\rowlabel{Perpetual incr. within-\ensuremath{R^2} (\%)} & \regstat{31}{306} & \regstat{23}{018} & \regstat{17}{691} & \regstat{8}{459} & \regstat{7}{121} & \regstat{4}{287}\\

\addlinespace[2pt]
\rowlabel{Common daytime sample} &  &  & \regpair{\regcoef{0}{608}{***}}{\regse{0}{096}} & \regpair{\regcoef{0}{666}{***}}{\regse{0}{157}} & \regpair{\regcoef{0}{627}{***}}{\regse{0}{167}} & \regpair{\regcoef{0}{695}{***}}{\regse{0}{182}}\\
\rowlabel{Perpetual incr. within-\ensuremath{R^2} (\%)} &  &  & \regstat{18}{533} & \regstat{9}{634} & \regstat{7}{893} & \regstat{3}{442}\\

\midrule\end{tabular}
\endgroup

\par\nointerlineskip

\begingroup
\linespread{1.0}\fontsize{9.5}{11}\selectfont
\setlength{\tabcolsep}{2.0pt}
\renewcommand{\arraystretch}{1.0}

\begin{tabular}{@{}L{0.38\linewidth}*{6}{C{\dimexpr(\linewidth-0.38\linewidth-12\tabcolsep)/6\relax}}@{}}
\multicolumn{7}{@{}p{\linewidth}@{}}{\bfseries Panel B. Cumulative return regressions using all valid observations at each time}\\[2pt]
 & \multicolumn{2}{C{\dimexpr ((\linewidth-0.38\linewidth-12\tabcolsep)/6)*2+2\tabcolsep\relax}}{Overnight trading} & \multicolumn{2}{C{\dimexpr ((\linewidth-0.38\linewidth-12\tabcolsep)/6)*2+2\tabcolsep\relax}}{Premarket trading} & \multicolumn{2}{C{\dimexpr ((\linewidth-0.38\linewidth-12\tabcolsep)/6)*2+2\tabcolsep\relax}@{}}{Regular trading}\\
\cmidrule(lr){2-3}
\cmidrule(lr){4-5}
\cmidrule(lr){6-7}
Variable & \grouphead{Overnight\\+5 min} & \grouphead{Overnight\\+30 min} & 04:05 & 09:25 & \grouphead{Regular\\open} & \grouphead{Reopening\\day close}\\
\midrule
\rowlabel{Weekend perpetual return} & \regpair{\regcoef{0}{619}{***}}{\regse{0}{054}} & \regpair{\regcoef{0}{581}{***}}{\regse{0}{069}} & \regpair{\regcoef{0}{602}{***}}{\regse{0}{086}} & \regpair{\regcoef{0}{643}{***}}{\regse{0}{159}} & \regpair{\regcoef{0}{592}{***}}{\regse{0}{166}} & \regpair{\regcoef{0}{682}{***}}{\regse{0}{166}}\\

\addlinespace[2pt]
\rowlabel{Stock-weeks} & 2,859 & 2,252 & 3,945 & 3,708 & 4,060 & 4,060\\
\rowlabel{Stocks} & 225 & 204 & 299 & 284 & 319 & 319\\
\rowlabel{Within-\ensuremath{R^2}} & \regstat{0}{312} & \regstat{0}{249} & \regstat{0}{179} & \regstat{0}{101} & \regstat{0}{080} & \regstat{0}{042}\\
\rowlabel{Perpetual incr. within-\ensuremath{R^2} (\%)} & \regstat{30}{627} & \regstat{23}{618} & \regstat{17}{290} & \regstat{8}{557} & \regstat{6}{994} & \regstat{3}{122}\\

\midrule\end{tabular}
\endgroup

\par\nointerlineskip

\begingroup
\linespread{1.0}\fontsize{9.5}{11}\selectfont
\setlength{\tabcolsep}{2.2pt}
\renewcommand{\arraystretch}{1.0}

\begin{tabular}{@{}L{0.38\linewidth}*{4}{C{\dimexpr(\linewidth-0.38\linewidth-8\tabcolsep)/4\relax}}@{}}
\multicolumn{5}{@{}p{\linewidth}@{}}{\bfseries Panel C. Interval stock return regressions}\\[2pt]
Variable & \grouphead{Overnight +5 min\\to 04:05} & \grouphead{04:05\\to 09:25} & \grouphead{09:25\\to regular open} & \grouphead{Overnight +5 min\\to regular open}\\
\midrule
\rowlabel{Weekend perpetual return} & \regpair{\regcoef{-0}{055}{}}{\regse{0}{071}} & \regpair{\regcoef{0}{022}{}}{\regse{0}{073}} & \regpair{\regcoef{-0}{030}{}}{\regse{0}{021}} & \regpair{\regcoef{-0}{063}{}}{\regse{0}{143}}\\

\addlinespace[2pt]
\rowlabel{Stock-weeks} & 1,408 & 1,408 & 1,408 & 1,408\\
\rowlabel{Stocks} & 118 & 118 & 118 & 118\\
\rowlabel{Within-\ensuremath{R^2}} & \regstat{0}{012} & \regstat{0}{019} & \regstat{0}{017} & \regstat{0}{016}\\
\rowlabel{Perpetual incr. within-\ensuremath{R^2} (\%)} & \regstat{0}{336} & \regstat{0}{023} & \regstat{0}{280} & \regstat{0}{116}\\

\bottomrule\end{tabular}
\endgroup

\begin{exhibitnotes}
All models control for historical stock characteristics and include stock and reopening-date fixed effects.
Panels A and B use cumulative stock returns at the indicated times (\%) as dependent variables
and report the coefficients on weekend perpetual returns.
In Panel A, the common overnight and daytime sample covers overnight, premarket, and regular trading and contains 1,408 stock-weeks for 118 stocks. The common daytime sample covers premarket and regular trading and contains 2,250 stock-weeks for 182 stocks. Each common sample requires valid prices at all observation times used in the corresponding analysis.
Panel B uses all valid observations at each time.
Panel C uses the common overnight and daytime sample
and stock returns over the indicated intervals (\%) as dependent variables.
\incrnote{}\regnote{}
\end{exhibitnotes}
\end{table}

\subsection{Stock Price Updates and the Incremental Information in Perpetual Returns}

Keeping the cumulative opening return as the dependent variable, we examine how the incremental explanatory power of weekend perpetual returns changes as stock prices update. Panel A of \tabref{tab:6} uses the same 3,708 stock-weeks in all specifications. Conditional on historical stock characteristics, adding weekend perpetual returns raises within-$R^2$ by 7.162 percentage points. After controlling for the cumulative stock return at 09:25, the increment falls to 0.024 percentage points and the perpetual return coefficient falls from 0.608 to $-0.037$, with the latter significant at the 10\% level.

The decline in incremental explanatory power is already evident early in overnight trading. In the common overnight and daytime sample of 1,408 stock-weeks, controlling for cumulative stock returns five minutes after the overnight open reduces incremental within-$R^2$ from 7.121 to 0.070 percentage points. When cumulative stock returns at each subsequent observation time are included separately, incremental within-$R^2$ remains below 0.1 percentage points and the perpetual return coefficients are insignificant (\tabref{tab:6}, Panel B).

Equivalence tests assess the economic magnitude of the remaining opening effect, measured in basis points per one-percentage-point higher weekend perpetual return. Five minutes after the overnight open, the estimate remains imprecise and does not pass the equivalence test with bounds of $\pm20$ basis points. Controlling for the 04:05 cumulative stock return yields equivalence within $\pm20$ basis points. At 09:25, the remaining opening effect is $-3.69$ basis points and passes the test with bounds of $\pm10$ basis points (\tabref{tab:6}, Panel C). Confidence intervals from the wild cluster bootstrap at the calendar-week level support the same equivalence conclusions (\apptabref{tab:B3}, Panel A).

Incremental within-$R^2$ also remains below 0.1 percentage points in alternative valid samples after controlling for cumulative stock returns during overnight or premarket trading (\apptabref{tab:B3}, Panel B). For cumulative opening returns, incremental explanatory power is therefore low once early overnight stock returns are included, and the remaining opening effect passes the $\pm10$-basis-point equivalence test conditional on the 09:25 cumulative stock return. Results for alternative price measures and return windows differ: the conditional perpetual return coefficients are $-0.246$ for the 09:25--09:35 quote-midpoint return and $-0.081$ for the cumulative return measured using VWAP over the first five minutes of the regular session, both significant at the 5\% level (\apptabref{tab:B3}, Panel C).

\begin{table}[H]
\centering\setlength{\parskip}{0pt}
\caption{Stock price updates and the incremental information in perpetual returns}\label{tab:6}
\begingroup\linespread{1.0}\fontsize{9}{10.4}\selectfont\setlength{\tabcolsep}{2.2pt}\renewcommand{\arraystretch}{1.0}
\begin{tabular}{@{}L{0.36\linewidth}*{4}{C{\dimexpr(\linewidth-0.36\linewidth-8\tabcolsep)/4\relax}}@{}}\toprule
\multicolumn{5}{@{}p{\linewidth}@{}}{\bfseries Panel A. Incremental explanatory power before and after controlling for premarket returns}\\[2pt]
Variable & \grouphead{Historical\\characteristics} & \grouphead{Add perpetual\\return} & \grouphead{Add 09:25\\stock return} & \grouphead{Add both\\returns}\\\midrule
\rowlabel{Weekend perpetual return} &  & \regpair{\regcoef{0}{608}{***}}{\regse{0}{169}} &  & \regpair{\regcoef{-0}{037}{*}}{\regse{0}{020}}\\
\rowlabel{09:25 cumulative stock return} &  &  & \regpair{\regcoef{0}{997}{***}}{\regse{0}{013}} & \regpair{\regcoef{1}{002}{***}}{\regse{0}{013}}\\
\rowlabel{Perpetual incr. within-\ensuremath{R^2} (\%)} &  & \regstat{7}{162} &  & \regstat{0}{024}\\
\rowlabel{Perpetual partial \ensuremath{R^2} (\%)} &  & \regstat{7}{240} &  & \regstat{0}{339}\\
\addlinespace[2pt]
\rowlabel{Stock-weeks} & 3,708 & 3,708 & 3,708 & 3,708\\
\rowlabel{Stocks} & 284 & 284 & 284 & 284\\
\rowlabel{Within-\ensuremath{R^2}} & \regstat{0}{01080} & \regstat{0}{08242} & \regstat{0}{92874} & \regstat{0}{92898}\\
\midrule\end{tabular}\endgroup
\par\nointerlineskip
\begingroup\linespread{1.0}\fontsize{9}{10.4}\selectfont\setlength{\tabcolsep}{2.0pt}\renewcommand{\arraystretch}{1.0}
\begin{tabular}{@{}L{0.36\linewidth}*{6}{C{\dimexpr(\linewidth-0.36\linewidth-12\tabcolsep)/6\relax}}@{}}
\multicolumn{7}{@{}p{\linewidth}@{}}{\bfseries Panel B. Conditioning on cumulative stock returns at different times}\\[2pt]
Variable & \grouphead{No stock\\return control} & \grouphead{Overnight\\+5 min} & \grouphead{Overnight\\+30 min} & 04:05 & 08:00 & 09:25\\\midrule
\rowlabel{Weekend perpetual return} & \regpair{\regcoef{0}{573}{***}}{\regse{0}{170}} & \regpair{\regcoef{-0}{069}{}}{\regse{0}{120}} & \regpair{\regcoef{0}{013}{}}{\regse{0}{118}} & \regpair{\regcoef{-0}{067}{}}{\regse{0}{060}} & \regpair{\regcoef{-0}{019}{}}{\regse{0}{051}} & \regpair{\regcoef{-0}{033}{}}{\regse{0}{024}}\\
\rowlabel{Overnight +5 min cumulative return} &  & \regpair{\regcoef{1}{009}{***}}{\regse{0}{107}} &  &  &  & \\
\rowlabel{Overnight +30 min cumulative return} &  &  & \regpair{\regcoef{0}{968}{***}}{\regse{0}{091}} &  &  & \\
\rowlabel{04:05 cumulative return} &  &  &  & \regpair{\regcoef{1}{102}{***}}{\regse{0}{065}} &  & \\
\rowlabel{08:00 cumulative return} &  &  &  &  & \regpair{\regcoef{1}{006}{***}}{\regse{0}{033}} & \\
\rowlabel{09:25 cumulative return} &  &  &  &  &  & \regpair{\regcoef{1}{005}{***}}{\regse{0}{017}}\\
\rowlabel{Within-\ensuremath{R^2}} & \regstat{0}{07934} & \regstat{0}{27310} & \regstat{0}{30357} & \regstat{0}{49274} & \regstat{0}{70217} & \regstat{0}{93220}\\
\rowlabel{Perpetual incr. within-\ensuremath{R^2} (\%)} & \regstat{7}{121} & \regstat{0}{070} & \regstat{0}{003} & \regstat{0}{081} & \regstat{0}{007} & \regstat{0}{021}\\
\midrule\end{tabular}\endgroup
\par\nointerlineskip
\begingroup\linespread{1.0}\fontsize{9}{10.4}\selectfont\setlength{\tabcolsep}{2.2pt}\renewcommand{\arraystretch}{1.0}
\begin{tabular}{@{}L{\dimexpr 0.290\linewidth-2.900\tabcolsep\relax}C{\dimexpr 0.130\linewidth-1.300\tabcolsep\relax}C{\dimexpr 0.230\linewidth-2.300\tabcolsep\relax}C{\dimexpr 0.115\linewidth-1.150\tabcolsep\relax}C{\dimexpr 0.115\linewidth-1.150\tabcolsep\relax}C{\dimexpr 0.120\linewidth-1.200\tabcolsep\relax}@{}}
\multicolumn{6}{@{}p{\linewidth}@{}}{\bfseries Panel C. Equivalence tests of the remaining opening effect}\\[2pt]
\grouphead{Stock return\\control time} & \grouphead{Remaining\\effect (bp)} & \grouphead{90\% confidence\\interval} & \ensuremath{\pm}5 bp & \ensuremath{\pm}10 bp & \ensuremath{\pm}20 bp\\\midrule
\rowlabel{Overnight +5 min} & \tabnum{-6.86} & \tabnum{[-27.06, 13.35]} & Fail & Fail & Fail\\
\rowlabel{04:05} & \tabnum{-1.78} & \tabnum{[-14.63, 11.06]} & Fail & Fail & Pass\\
\rowlabel{09:25} & \tabnum{-3.69} & \tabnum{[-6.98, -0.40]} & Fail & Pass & Pass\\
\bottomrule\end{tabular}\endgroup
\begin{exhibitnotes}
All models use the cumulative opening return as the dependent variable, control for historical stock characteristics, and include stock and reopening-date fixed effects. Panel A uses the same 3,708 stock-weeks to compare the increase in within-$R^2$ from adding weekend perpetual returns, first with historical stock characteristics alone and then also controlling for the 09:25 cumulative stock return. Panel B uses 1,408 stock-weeks from the common overnight and daytime sample (118 stocks) and controls separately for cumulative stock returns at each indicated time. Panel C reports the remaining opening effect, in basis points per one-percentage-point higher weekend perpetual return, and 90\% confidence intervals based on two-way clustering. The sample sizes for overnight +5 minutes, 04:05, and 09:25 are 1,408, 2,250, and 3,708 stock-weeks, respectively. Equivalence tests use a 5\% significance level. \incrnote{} Partial $R^2$ is reported in \%. \regnote{}
\end{exhibitnotes}
\end{table}

\subsection{Short-Run Persistence of Stock Price Adjustments}

\tabref{tab:7} examines whether stock price adjustments associated with weekend perpetual returns persist after the regular-session open. In a common sample of 3,949 stock-weeks, the perpetual return coefficients for cumulative stock returns at the regular-session open, the close of the reopening day, and the close of the next trading day are 0.477, 0.623, and 0.718, respectively, all significant at the 1\% level. By contrast, the coefficients for stock returns measured from the regular-session open are insignificant at every reported horizon. The positive association with cumulative stock returns thus extends through the close of the next trading day. For longer windows ending at the close of the third and fifth trading days after reopening, the point estimates are 0.336 and 0.208, respectively. The former is significant at the 10\% level and the latter is not (\apptabref{tab:B4}, Panel A, columns (1) and (2)).

In the baseline regressions with four weekend return intervals, the coefficient gap between the final and first intervals narrows between the regular-session open and the close of the reopening day. Panel B of \apptabref{tab:B4} shows that the gap falls from 0.652 for cumulative opening returns to 0.161 for cumulative closing returns. The corresponding coefficient difference for open-to-close returns is $-0.491$, significant at the 5\% level. It remains significantly negative after controlling for ES market exposure (\apptabref{tab:B4}, Panel A, column (4)).

\begin{table}[H]
\centering\setlength{\parskip}{0pt}
\caption{Weekend perpetual returns and the short-run persistence of stock price adjustments}\label{tab:7}

\begingroup
\linespread{1.0}\fontsize{9.5}{11}\selectfont
\setlength{\tabcolsep}{2.0pt}
\renewcommand{\arraystretch}{1.0}

\begin{tabular}{@{}L{0.36\linewidth}*{6}{C{\dimexpr(\linewidth-0.36\linewidth-12\tabcolsep)/6\relax}}@{}}
\toprule
\multicolumn{7}{@{}p{\linewidth}@{}}{\bfseries Panel A. Cumulative stock returns}\\[2pt]
Variable & \grouphead{Regular\\open} & \grouphead{Open\\+5 min} & \grouphead{Open\\+15 min} & \grouphead{Open\\+60 min} & \grouphead{Reopening\\day close} & \grouphead{Next trading\\day close}\\
\midrule
\rowlabel{Weekend perpetual return} & \regpair{\regcoef{0}{477}{***}}{\regse{0}{136}} & \regpair{\regcoef{0}{472}{***}}{\regse{0}{149}} & \regpair{\regcoef{0}{498}{***}}{\regse{0}{122}} & \regpair{\regcoef{0}{498}{***}}{\regse{0}{126}} & \regpair{\regcoef{0}{623}{***}}{\regse{0}{166}} & \regpair{\regcoef{0}{718}{***}}{\regse{0}{225}}\\
\rowlabel{Within-\ensuremath{R^2}} & \regstat{0}{063} & \regstat{0}{043} & \regstat{0}{039} & \regstat{0}{033} & \regstat{0}{035} & \regstat{0}{041}\\
\rowlabel{Perpetual incr. within-\ensuremath{R^2} (\%)} & \regstat{5}{550} & \regstat{3}{310} & \regstat{3}{197} & \regstat{2}{541} & \regstat{2}{645} & \regstat{1}{994}\\

\midrule
\multicolumn{7}{@{}p{\linewidth}@{}}{\bfseries Panel B. Stock returns measured from the regular-session open}\\[2pt]
\rowlabel{Weekend perpetual return} & \textemdash{} & \regpair{\regcoef{-0}{006}{}}{\regse{0}{034}} & \regpair{\regcoef{0}{021}{}}{\regse{0}{042}} & \regpair{\regcoef{0}{021}{}}{\regse{0}{100}} & \regpair{\regcoef{0}{146}{}}{\regse{0}{127}} & \regpair{\regcoef{0}{241}{}}{\regse{0}{214}}\\
\rowlabel{Within-\ensuremath{R^2}} & \textemdash{} & \regstat{0}{015} & \regstat{0}{012} & \regstat{0}{010} & \regstat{0}{009} & \regstat{0}{028}\\
\rowlabel{Perpetual incr. within-\ensuremath{R^2} (\%)} & \textemdash{} & \regstat{0}{001} & \regstat{0}{013} & \regstat{0}{008} & \regstat{0}{212} & \regstat{0}{263}\\

\bottomrule\end{tabular}
\endgroup

\begin{exhibitnotes}
All models use the same 3,949 stock-weeks for 319 stocks,
control for historical stock characteristics, and include stock and reopening-date fixed effects.
Panel A uses cumulative stock returns from Friday 20:00 to the indicated time (\%) as dependent variables.
Panel B uses stock returns from the regular-session open to the indicated time (\%).
\incrnote{}\regnote{}
\end{exhibitnotes}
\end{table}

\FloatBarrier
\section{Microstructure Evidence and Heterogeneity}\label{sec:microstructure}

\subsection{Price Updates Early in the Reopening Process}

Early stock price responses can appear both in the first overnight transaction price and in subsequent updates within the same minute. We measure the first transaction price using the opening price of the first minute with valid overnight trades and use that minute's opening and closing prices to examine these two components. In the 3,603 stock-weeks in Panel A of \tabref{tab:8}, the coefficients on weekend perpetual returns are 0.489 and 0.605 for cumulative stock returns measured at the minute's opening and closing prices, respectively, and 0.115 for the within-minute stock return. All three are significant at the 1\% level. A one-percentage-point higher weekend perpetual return is associated with an 11.5-basis-point higher within-minute stock return. This coefficient remains significantly positive when we restrict the sample to minutes with at least 2, 5, or 10 trades.

The perpetual return coefficient is also significantly positive when within-minute stock returns are measured using quote midpoints within the minute, quote midpoints matched to transaction times, VWAPs for the first and second 30-second intervals, or transaction prices excluding odd-lot trades (\tabref{tab:8}, Panel B). In the respective common samples, the coefficients for the 30-second VWAP measure and the measure excluding odd-lot trades are 0.088 and 0.146. Both are significantly smaller than the corresponding coefficients based on the original transaction price measure.

In the common ES sample for each price measure, neither quote-midpoint coefficient is significant even before ES market exposure is included. Adding ES market exposure also renders the 30-second VWAP coefficient insignificant, whereas the coefficient for returns excluding odd-lot trades remains significant at the 5\% level (\apptabref{tab:C1}).

Price updates from the opening to the closing price of the first minute with valid overnight trades contain incremental information about cumulative opening returns. In Panel C of \tabref{tab:8}, replacing the cumulative stock return at the minute's opening price with the return at its closing price reduces the conditional coefficient on perpetual returns from 0.320 to 0.156. The difference is significant at the 5\% level, and the incremental within-$R^2$ falls from 1.548 to 0.328 percentage points. When weekend perpetual returns and the cumulative stock return at the minute's opening price are both included, the coefficient on the within-minute stock return is 0.706, significant at the 1\% level.

\begin{table}[H]
\centering\setlength{\parskip}{0pt}
\caption{Early price updates after reopening: evidence from trades and quotes}\label{tab:8}

\begingroup
\linespread{1.0}\fontsize{8.5}{9.8}\selectfont
\setlength{\tabcolsep}{2.0pt}
\renewcommand{\arraystretch}{1.0}

\begin{tabular}{@{}L{0.40\linewidth}*{6}{C{\dimexpr(\linewidth-0.40\linewidth-12\tabcolsep)/6\relax}}@{}}
\toprule
\multicolumn{7}{@{}p{\linewidth}@{}}{\bfseries Panel A. First-minute price updates and minimum trade counts}\\[0.5pt]
 & \multicolumn{3}{C{\dimexpr ((\linewidth-0.40\linewidth-12\tabcolsep)/6)*3+4\tabcolsep\relax}}{First minute with valid trades} & \multicolumn{3}{C{\dimexpr ((\linewidth-0.40\linewidth-12\tabcolsep)/6)*3+4\tabcolsep\relax}@{}}{Within-minute returns by minimum trade count}\\
\cmidrule(lr){2-4}
\cmidrule(lr){5-7}
Variable & Cum. return at minute open & Cum. return at minute close & Within-minute return & At least 2 trades & At least 5 trades & At least 10 trades\\
\midrule
\rowlabel{Weekend perpetual return} & \regpair{\regcoef{0}{489}{***}}{\regse{0}{058}} & \regpair{\regcoef{0}{605}{***}}{\regse{0}{053}} & \regpair{\regcoef{0}{115}{***}}{\regse{0}{034}} & \regpair{\regcoef{0}{110}{***}}{\regse{0}{033}} & \regpair{\regcoef{0}{119}{***}}{\regse{0}{035}} & \regpair{\regcoef{0}{124}{***}}{\regse{0}{043}}\\

\addlinespace[0.5pt]
\rowlabel{Stock-weeks} & 3,603 & 3,603 & 3,603 & 3,476 & 3,005 & 2,432\\
\rowlabel{Stocks} & 266 & 266 & 266 & 259 & 231 & 197\\
\rowlabel{Within-\ensuremath{R^2}} & \regstat{0}{255} & \regstat{0}{335} & \regstat{0}{036} & \regstat{0}{033} & \regstat{0}{037} & \regstat{0}{038}\\
\rowlabel{Perpetual incr. within-\ensuremath{R^2} (\%)} & \regstat{24}{044} & \regstat{32}{336} & \regstat{3}{463} & \regstat{3}{138} & \regstat{3}{464} & \regstat{3}{348}\\

\midrule\end{tabular}
\endgroup

\par\nointerlineskip

\begingroup
\linespread{1.0}\fontsize{8.5}{9.8}\selectfont
\setlength{\tabcolsep}{2.2pt}
\renewcommand{\arraystretch}{1.0}

\begin{tabular}{@{}L{0.40\linewidth}*{4}{C{\dimexpr(\linewidth-0.40\linewidth-8\tabcolsep)/4\relax}}@{}}
\multicolumn{5}{@{}p{\linewidth}@{}}{\bfseries Panel B. Within-minute returns using alternative price measures}\\[0.5pt]
Variable & Within-minute quote midpoint & Quote midpoint matched to trades & 30-second VWAP & Excluding odd-lot trades\\
\midrule
\rowlabel{Original transaction price measure} & \regpair{\regcoef{0}{245}{***}}{\regse{0}{069}} & \regpair{\regcoef{0}{210}{***}}{\regse{0}{071}} & \regpair{\regcoef{0}{324}{***}}{\regse{0}{094}} & \regpair{\regcoef{0}{243}{***}}{\regse{0}{067}}\\
\rowlabel{Alternative price measure} & \regpair{\regcoef{0}{200}{**}}{\regse{0}{098}} & \regpair{\regcoef{0}{373}{**}}{\regse{0}{147}} & \regpair{\regcoef{0}{088}{***}}{\regse{0}{030}} & \regpair{\regcoef{0}{146}{***}}{\regse{0}{046}}\\
\rowlabel{Coefficient difference (original minus alternative)} & \regpair{\regcoef{0}{045}{}}{\regse{0}{131}} & \regpair{\regcoef{-0}{164}{}}{\regse{0}{105}} & \regpair{\regcoef{0}{236}{**}}{\regse{0}{095}} & \regpair{\regcoef{0}{097}{***}}{\regse{0}{029}}\\

\addlinespace[0.5pt]
\rowlabel{Stock-weeks} & 710 & 491 & 452 & 718\\
\rowlabel{Stocks} & 204 & 183 & 144 & 207\\
\rowlabel{Coefficient difference: $p$-value} & \regstat{0}{7310} & \regstat{0}{1282} & \regstat{0}{0177} & \regstat{0}{0018}\\

\midrule\end{tabular}
\endgroup

\par\nointerlineskip

\begingroup
\linespread{1.0}\fontsize{8.5}{9.8}\selectfont
\setlength{\tabcolsep}{2.2pt}
\renewcommand{\arraystretch}{1.0}

\begin{tabular}{@{}L{0.40\linewidth}*{4}{C{\dimexpr(\linewidth-0.40\linewidth-8\tabcolsep)/4\relax}}@{}}
\multicolumn{5}{@{}p{\linewidth}@{}}{\bfseries Panel C. First-minute price updates and incremental explanatory power of perpetual returns}\\[0.5pt]
Variable & No first-minute return control & Control: cum. return at minute open & Control: cum. return at minute close & Cum. return at minute open and within-minute return\\
\midrule
\rowlabel{Weekend perpetual return} & \regpair{\regcoef{0}{603}{***}}{\regse{0}{165}} & \regpair{\regcoef{0}{320}{**}}{\regse{0}{139}} & \regpair{\regcoef{0}{156}{}}{\regse{0}{156}} & \regpair{\regcoef{0}{154}{}}{\regse{0}{151}}\\
\rowlabel{Cum. return at first-minute open} &  & \regpair{\regcoef{0}{579}{***}}{\regse{0}{128}} &  & \regpair{\regcoef{0}{751}{***}}{\regse{0}{121}}\\
\rowlabel{Cum. return at first-minute close} &  &  & \regpair{\regcoef{0}{739}{***}}{\regse{0}{116}} & \\
\rowlabel{Within-minute stock return} &  &  &  & \regpair{\regcoef{0}{706}{***}}{\regse{0}{221}}\\
\rowlabel{Historical stock characteristics} & YES & YES & YES & YES\\
\rowlabel{Stock fixed effects} & YES & YES & YES & YES\\
\rowlabel{Reopening-date fixed effects} & YES & YES & YES & YES\\

\addlinespace[0.5pt]
\rowlabel{Stock-weeks} & 3,603 & 3,603 & 3,603 & 3,603\\
\rowlabel{Stocks} & 266 & 266 & 266 & 266\\
\rowlabel{Within-\ensuremath{R^2}} & \regstat{0}{083} & \regstat{0}{133} & \regstat{0}{165} & \regstat{0}{166}\\
\rowlabel{Perpetual incr. within-\ensuremath{R^2} (\%)} & \regstat{7}{277} & \regstat{1}{548} & \regstat{0}{328} & \regstat{0}{320}\\

\bottomrule\end{tabular}
\endgroup

\begin{exhibitnotes}
All models include historical stock characteristics, stock fixed effects, and reopening-date fixed effects. Panel A uses cumulative stock returns at the opening and closing prices of the first minute with valid overnight trades, as well as the within-minute stock return. Panel B compares coefficients on weekend perpetual returns using the original transaction price measure and each alternative measure in a common sample within each column. Regressions of return differences test the differences in coefficients. Quotes are matched at the transaction time or within the preceding 5 seconds. VWAP returns compare volume-weighted prices in the first and second 30-second intervals. Odd-lot trades are identified by trade condition I. Panel C uses the cumulative opening return as the dependent variable and includes weekend perpetual returns and the indicated first-minute return controls. The difference between the perpetual return coefficients when controlling for the cumulative return at the minute's closing rather than opening price is $-0.164$ (two-way clustered standard error 0.078; $p=0.043$). The 95\% confidence interval from a paired bootstrap at the calendar-week level is [$-0.302$, $-0.036$].
\incrnote{}\regnote{}
\end{exhibitnotes}
\end{table}

\subsection{Cross-Venue Return Dispersion and Early Price Responses}

Trading conditions differ across perpetual venues, and weekend returns on contracts linked to the same stock may differ across venues. We examine whether cross-venue return dispersion is associated with the sensitivity of early stock returns to weekend perpetual returns. Let $D_{iw}$ denote the sample standard deviation of weekend perpetual returns across venues for a given stock-week, requiring at least two venues with valid returns. We center dispersion as $D^c_{iw}=D_{iw}-\bar D$, where $\bar D$ is its mean in the relevant estimation sample, and estimate
\begin{equation}\label{eq:heterogeneity}
\begin{gathered}
 R_{iw}(t)=\alpha_i+\delta_w+\beta W_{iw}
 +\eta D^c_{iw}+\lambda W_{iw}D^c_{iw}\\
 {}+\gamma'H_{iw}+\phi'Z_{iw}
 +\psi'(W_{iw}Z_{iw})
 +a|W_{iw}|+bW_{iw}|W_{iw}|+e_{iwt}.
\end{gathered}
\end{equation}
Here, $Z_{iw}$ includes the number of venues with valid returns, historical perpetual trading activity, historical overnight stock trading activity, and the perpetual price observation lag at the Sunday cutoff. The interaction coefficient $\lambda$ measures how the conditional relation between perpetual and stock returns varies with cross-venue return dispersion. The model also includes nonlinear terms in perpetual returns and interactions between perpetual returns and the other trading characteristics. \apptabref{tab:A4} defines these variables.

In the 2,109 stock-weeks with the required venue prices, early stock prices, and trading characteristics, the interaction coefficient in the full specification is 0.215, significant at the 1\% level (\tabref{tab:9}, Panel A). The corresponding estimates in the common overnight and daytime sample and the common daytime sample are 0.220 and 0.219, both significant at the 1\% level. The interaction remains significantly positive when dispersion is measured by the median absolute deviation or interquartile range, when the price observation window is restricted to five minutes, or when extreme return observations are excluded (\apptabref{tab:C2}).

To identify when this interaction emerges, we distinguish the first overnight transaction price from the price update between that transaction and five minutes after the overnight open (\tabref{tab:9}, Panel B). The interaction coefficient is $-0.030$ and insignificant for the cumulative return at the first transaction price, but 0.245 and significant at the 1\% level for the subsequent interval return. The latter exceeds the former by 0.275, a difference significant at the 5\% level. The interaction is insignificant for the stock return from five minutes after the overnight open to the regular-session open, but remains significantly positive for the cumulative opening return.

Panel C of \tabref{tab:9} reports the average marginal effect of perpetual returns at different levels of dispersion. Evaluated at the 25th and 75th percentiles of cross-venue return dispersion, the average marginal effects are 0.528 and 0.557. The difference is significant at the 1\% level and corresponds to approximately 2.9 basis points in early overnight stock returns per one-percentage-point higher weekend perpetual return.

\begin{table}[H]
\centering\setlength{\parskip}{0pt}
\caption{Cross-venue return dispersion and stock price responses}\label{tab:9}

\begingroup
\linespread{1.0}\fontsize{9}{10.4}\selectfont
\setlength{\tabcolsep}{2.0pt}
\renewcommand{\arraystretch}{1.0}

\begin{tabular}{@{}L{0.40\linewidth}*{6}{C{\dimexpr(\linewidth-0.40\linewidth-12\tabcolsep)/6\relax}}@{}}
\toprule
\multicolumn{7}{@{}p{\linewidth}@{}}{\bfseries Panel A. Return dispersion and cumulative returns five minutes after the overnight open}\\[0.5pt]
Variable & (1) & (2) & (3) & (4) & (5) & (6)\\
\midrule
\rowlabel{Weekend perpetual return} & \regpair{\regcoef{0}{674}{***}}{\regse{0}{049}} & \regpair{\regcoef{0}{672}{***}}{\regse{0}{048}} & \regpair{\regcoef{0}{645}{***}}{\regse{0}{046}} & \regpair{\regcoef{0}{546}{***}}{\regse{0}{060}} & \regpair{\regcoef{0}{547}{***}}{\regse{0}{060}} & \regpair{\regcoef{0}{547}{***}}{\regse{0}{147}}\\
\rowlabel{Return dispersion (centered)} &  & \regpair{\regcoef{-0}{229}{}}{\regse{0}{252}} & \regpair{\regcoef{0}{044}{}}{\regse{0}{213}} & \regpair{\regcoef{0}{136}{}}{\regse{0}{189}} & \regpair{\regcoef{0}{136}{}}{\regse{0}{189}} & \regpair{\regcoef{0}{138}{}}{\regse{0}{189}}\\
\rowlabel{Perpetual return \ensuremath{\times} dispersion} &  &  & \regpair{\regcoef{0}{254}{***}}{\regse{0}{046}} & \regpair{\regcoef{0}{206}{***}}{\regse{0}{049}} & \regpair{\regcoef{0}{206}{***}}{\regse{0}{049}} & \regpair{\regcoef{0}{215}{***}}{\regse{0}{048}}\\

\addlinespace[0.5pt]
\rowlabel{Nonlinear perpetual return terms} & NO & NO & NO & YES & YES & YES\\
\rowlabel{Other trading characteristics} & NO & NO & NO & NO & YES & YES\\
\rowlabel{Perpetual return \ensuremath{\times} other trading characteristics} & NO & NO & NO & NO & NO & YES\\
\rowlabel{Historical stock characteristics} & YES & YES & YES & YES & YES & YES\\
\rowlabel{Stock fixed effects} & YES & YES & YES & YES & YES & YES\\
\rowlabel{Reopening-date fixed effects} & YES & YES & YES & YES & YES & YES\\

\addlinespace[0.5pt]
\rowlabel{Stock-weeks} & 2,109 & 2,109 & 2,109 & 2,109 & 2,109 & 2,109\\
\rowlabel{Stocks} & 168 & 168 & 168 & 168 & 168 & 168\\
\rowlabel{Within-\ensuremath{R^2}} & \regstat{0}{447} & \regstat{0}{449} & \regstat{0}{470} & \regstat{0}{484} & \regstat{0}{484} & \regstat{0}{488}\\

\midrule\end{tabular}
\endgroup

\par\nointerlineskip

\begingroup
\linespread{1.0}\fontsize{9}{10.4}\selectfont
\setlength{\tabcolsep}{2.0pt}
\renewcommand{\arraystretch}{1.0}

\begin{tabular}{@{}L{0.40\linewidth}*{5}{C{\dimexpr(\linewidth-0.40\linewidth-10\tabcolsep)/5\relax}}@{}}
\multicolumn{6}{@{}p{\linewidth}@{}}{\bfseries Panel B. Return dispersion interactions across reopening stages}\\[0.5pt]
 & \multicolumn{1}{C{\dimexpr ((\linewidth-0.40\linewidth-10\tabcolsep)/5)*1+0\tabcolsep\relax}}{Cumulative return} & \multicolumn{1}{C{\dimexpr ((\linewidth-0.40\linewidth-10\tabcolsep)/5)*1+0\tabcolsep\relax}}{Interval return} & \multicolumn{1}{C{\dimexpr ((\linewidth-0.40\linewidth-10\tabcolsep)/5)*1+0\tabcolsep\relax}}{Cumulative return} & \multicolumn{1}{C{\dimexpr ((\linewidth-0.40\linewidth-10\tabcolsep)/5)*1+0\tabcolsep\relax}}{Interval return} & \multicolumn{1}{C{\dimexpr ((\linewidth-0.40\linewidth-10\tabcolsep)/5)*1+0\tabcolsep\relax}@{}}{Cumulative return}\\
\cmidrule(lr){2-2}
\cmidrule(lr){3-3}
\cmidrule(lr){4-4}
\cmidrule(lr){5-5}
\cmidrule(lr){6-6}
Variable & First overnight transaction price & First transaction to overnight +5 min & Overnight +5 min & Overnight +5 min to regular-session open & Regular-session open\\
\midrule
\rowlabel{Weekend perpetual return} & \regpair{\regcoef{0}{605}{**}}{\regse{0}{296}} & \regpair{\regcoef{-0}{058}{}}{\regse{0}{235}} & \regpair{\regcoef{0}{547}{***}}{\regse{0}{147}} & \regpair{\regcoef{1}{406}{***}}{\regse{0}{492}} & \regpair{\regcoef{1}{953}{***}}{\regse{0}{500}}\\
\rowlabel{Return dispersion (centered)} & \regpair{\regcoef{0}{045}{}}{\regse{0}{142}} & \regpair{\regcoef{0}{093}{}}{\regse{0}{185}} & \regpair{\regcoef{0}{138}{}}{\regse{0}{189}} & \regpair{\regcoef{-0}{301}{}}{\regse{0}{323}} & \regpair{\regcoef{-0}{163}{}}{\regse{0}{379}}\\
\rowlabel{Perpetual return \ensuremath{\times} dispersion} & \regpair{\regcoef{-0}{030}{}}{\regse{0}{059}} & \regpair{\regcoef{0}{245}{***}}{\regse{0}{069}} & \regpair{\regcoef{0}{215}{***}}{\regse{0}{048}} & \regpair{\regcoef{0}{141}{}}{\regse{0}{112}} & \regpair{\regcoef{0}{355}{***}}{\regse{0}{119}}\\

\addlinespace[0.5pt]
\rowlabel{Stock-weeks} & 2,109 & 2,109 & 2,109 & 2,109 & 2,109\\
\rowlabel{Stocks} & 168 & 168 & 168 & 168 & 168\\
\rowlabel{Within-\ensuremath{R^2}} & \regstat{0}{449} & \regstat{0}{118} & \regstat{0}{488} & \regstat{0}{068} & \regstat{0}{180}\\

\midrule\end{tabular}
\endgroup

\par\nointerlineskip

\begingroup
\linespread{1.0}\fontsize{9}{10.4}\selectfont
\setlength{\tabcolsep}{2.0pt}
\renewcommand{\arraystretch}{1.0}

\begin{tabular}{@{}L{0.40\linewidth}*{5}{C{\dimexpr(\linewidth-0.40\linewidth-10\tabcolsep)/5\relax}}@{}}
\multicolumn{6}{@{}p{\linewidth}@{}}{\bfseries Panel C. Average marginal effects at different levels of dispersion}\\[0.5pt]
Variable & P25 & P50 & P75 & P90 & P75\ensuremath{-}P25\\
\midrule
\rowlabel{Average marginal effect of perpetual returns} & \regpair{\regcoef{0}{528}{***}}{\regse{0}{047}} & \regpair{\regcoef{0}{538}{***}}{\regse{0}{048}} & \regpair{\regcoef{0}{557}{***}}{\regse{0}{050}} & \regpair{\regcoef{0}{594}{***}}{\regse{0}{055}} & \regpair{\regcoef{0}{029}{***}}{\regse{0}{006}}\\
\rowlabel{Return dispersion (\%)} & \regstat{0}{066} & \regstat{0}{114} & \regstat{0}{199} & \regstat{0}{372} & \\
\rowlabel{95\% confidence interval: Lower} & \regstat{0}{433} & \regstat{0}{441} & \regstat{0}{456} & \regstat{0}{483} & \regstat{0}{016}\\
\rowlabel{95\% confidence interval: Upper} & \regstat{0}{623} & \regstat{0}{636} & \regstat{0}{658} & \regstat{0}{705} & \regstat{0}{042}\\

\bottomrule\end{tabular}
\endgroup

\begin{exhibitnotes}
All columns in Panel A use the same 2,109 stock-weeks and the cumulative stock return five minutes after the overnight open (\%) as the dependent variable. Column (6) estimates the full model in \eqnref{eq:heterogeneity}. Panel B applies the full model to cumulative and interval stock returns at different reopening stages in the same 2,109 stock-weeks. In this sample, the first minute with valid overnight trades occurs within five minutes of the overnight open. The interaction coefficient for the return from the first transaction to five minutes after the overnight open exceeds the coefficient for the cumulative return at the first transaction by 0.275 ($p=0.026$). Panel C uses the estimates in Panel A, column (6), to calculate average marginal effects and their differences with dispersion set to its 25th, 50th, 75th, and 90th percentiles, holding other variables at their observed values. Standard errors and 95\% confidence intervals are calculated from the covariance matrix of the full model.
\regnote{}
\end{exhibitnotes}
\end{table}

\FloatBarrier
\section{Common Market Information and Robustness}\label{sec:robustness}

\subsection{Stock Index Futures Information and the Incremental Explanatory Power of Perpetual Returns}

To assess the relation between weekend perpetual returns and common market information, we control for each stock's exposure to ES returns using its historical market beta. We define ES market exposure as
\begin{equation}\label{eq:market_exposure}
 M_{iw}=\widehat\beta^{SPY}_{iw}R_w^{ES}.
\end{equation}
Here, $R_w^{ES}$ is the ES return from Friday 17:00 to Sunday 19:55, and $\widehat\beta^{SPY}_{iw}$ is estimated from returns on the stock and SPY, an ETF tracking the S\&P 500 Index, over the 60 trading days through Friday.

In the common sample of 3,091 stock-weeks with ES market exposure available, including this control reduces the coefficient on weekend perpetual returns in the cumulative opening return regression from 0.519 to 0.320 (\tabref{tab:10}, Panel A). The coefficient remains significant at the 5\% level, while the incremental within-$R^2$ falls from 4.788 to 1.652 percentage points. In the 2,205 stock-weeks with valid early overnight prices, the coefficient for cumulative stock returns five minutes after the overnight open declines from 0.608 to 0.533 and remains significant at the 1\% level.

We use the same ES market exposure measure, based on the ES return through Sunday 19:55, to compare the conditional relations between returns over different weekend intervals and cumulative opening returns. After controlling for ES market exposure, the coefficients on perpetual returns before and after ES reopens are 0.266 and 0.874, respectively, both significantly positive (\tabref{tab:10}, Panel B). In the regression including all four weekend return intervals, the difference between the final- and first-interval coefficients declines from 0.839 to 0.665 and remains significant at the 1\% level.

Among the 2,848 stock-weeks with both ES market exposure and cumulative stock returns at 09:25, the incremental within-$R^2$ of perpetual returns is 1.798 percentage points conditional on ES market exposure. Adding the 09:25 cumulative stock return reduces this increment to 0.005 percentage points. The remaining opening effect is $-1.86$ basis points per one-percentage-point higher weekend perpetual return and passes the equivalence test with bounds of $\pm5$ basis points (\tabref{tab:10}, Panel C). Confidence intervals from the wild cluster bootstrap at the calendar-week level support the same equivalence conclusion (\apptabref{tab:D1}).

\begin{table}[H]
\centering\setlength{\parskip}{0pt}
\caption{ES market exposure and the incremental explanatory power of perpetual returns}\label{tab:10}

\begingroup
\linespread{1.0}\fontsize{8.5}{9.8}\selectfont
\setlength{\tabcolsep}{2.2pt}
\renewcommand{\arraystretch}{1.0}

\begin{tabular}{@{}L{0.37\linewidth}*{4}{C{\dimexpr(\linewidth-0.37\linewidth-8\tabcolsep)/4\relax}}@{}}
\toprule
\multicolumn{5}{@{}p{\linewidth}@{}}{\bfseries Panel A. Stock return regressions with and without ES market exposure}\\[0.5pt]
 & \multicolumn{2}{C{\dimexpr ((\linewidth-0.37\linewidth-8\tabcolsep)/4)*2+2\tabcolsep\relax}}{Regular-session open} & \multicolumn{2}{C{\dimexpr ((\linewidth-0.37\linewidth-8\tabcolsep)/4)*2+2\tabcolsep\relax}@{}}{Overnight +5 min}\\
\cmidrule(lr){2-3}
\cmidrule(lr){4-5}
Variable & (1) Without ES & (2) With ES & (3) Without ES & (4) With ES\\
\midrule
\rowlabel{Weekend perpetual return} & \regpair{\regcoef{0}{519}{***}}{\regse{0}{187}} & \regpair{\regcoef{0}{320}{**}}{\regse{0}{133}} & \regpair{\regcoef{0}{608}{***}}{\regse{0}{070}} & \regpair{\regcoef{0}{533}{***}}{\regse{0}{066}}\\
\rowlabel{ES market exposure} &  & \regpair{\regcoef{0}{822}{**}}{\regse{0}{366}} &  & \regpair{\regcoef{0}{413}{**}}{\regse{0}{153}}\\

\addlinespace[0.5pt]
\rowlabel{Stock-weeks} & 3,091 & 3,091 & 2,205 & 2,205\\
\rowlabel{Stocks} & 293 & 293 & 211 & 211\\
\rowlabel{Within-\ensuremath{R^2}} & \regstat{0}{065} & \regstat{0}{136} & \regstat{0}{254} & \regstat{0}{309}\\
\rowlabel{Perpetual incr. within-\ensuremath{R^2} (\%)} & \regstat{4}{788} & \regstat{1}{652} & \regstat{25}{034} & \regstat{17}{958}\\

\midrule\end{tabular}
\endgroup

\par\nointerlineskip

\begingroup
\linespread{1.0}\fontsize{8.5}{9.8}\selectfont
\setlength{\tabcolsep}{2.2pt}
\renewcommand{\arraystretch}{1.0}

\begin{tabular}{@{}L{0.37\linewidth}*{4}{C{\dimexpr(\linewidth-0.37\linewidth-8\tabcolsep)/4\relax}}@{}}
\multicolumn{5}{@{}p{\linewidth}@{}}{\bfseries Panel B. Weekend interval returns and ES market exposure}\\[0.5pt]
 & \multicolumn{2}{C{\dimexpr ((\linewidth-0.37\linewidth-8\tabcolsep)/4)*2+2\tabcolsep\relax}}{Four weekend return intervals} & \multicolumn{2}{C{\dimexpr ((\linewidth-0.37\linewidth-8\tabcolsep)/4)*2+2\tabcolsep\relax}@{}}{Two weekend return intervals}\\
\cmidrule(lr){2-3}
\cmidrule(lr){4-5}
Variable & (1) Without ES & (2) With ES & (3) Without ES & (4) With ES\\
\midrule
\rowlabel{Fri. 20:00 to Sat. 20:00} & \regpair{\regcoef{0}{358}{***}}{\regse{0}{114}} & \regpair{\regcoef{0}{227}{**}}{\regse{0}{088}} &  & \\
\rowlabel{Sat. 20:00 to Sun. 12:00} & \regpair{\regcoef{0}{437}{*}}{\regse{0}{216}} & \regpair{\regcoef{0}{236}{}}{\regse{0}{142}} &  & \\
\rowlabel{Sun. 12:00 to 18:00} & \regpair{\regcoef{0}{623}{**}}{\regse{0}{263}} & \regpair{\regcoef{0}{404}{*}}{\regse{0}{223}} &  & \\
\rowlabel{Sun. 18:00 to 19:55} & \regpair{\regcoef{1}{197}{***}}{\regse{0}{299}} & \regpair{\regcoef{0}{893}{***}}{\regse{0}{268}} &  & \\
\rowlabel{Pre-ES perpetual return} &  &  & \regpair{\regcoef{0}{434}{**}}{\regse{0}{162}} & \regpair{\regcoef{0}{266}{**}}{\regse{0}{118}}\\
\rowlabel{Post-ES perpetual return} &  &  & \regpair{\regcoef{1}{176}{***}}{\regse{0}{297}} & \regpair{\regcoef{0}{874}{***}}{\regse{0}{266}}\\
\rowlabel{ES market exposure} &  & \regpair{\regcoef{0}{756}{*}}{\regse{0}{376}} &  & \regpair{\regcoef{0}{762}{*}}{\regse{0}{377}}\\
\rowlabel{Final-minus-first coefficient} & \regpair{\regcoef{0}{839}{***}}{\regse{0}{233}} & \regpair{\regcoef{0}{665}{***}}{\regse{0}{237}} &  & \\
\rowlabel{Coefficient difference: $p$-value} & \regstat{0}{0011} & \regstat{0}{0085} &  & \\

\addlinespace[0.5pt]
\rowlabel{Stock-weeks} & 2,920 & 2,920 & 2,920 & 2,920\\
\rowlabel{Stocks} & 285 & 285 & 285 & 285\\
\rowlabel{Within-\ensuremath{R^2}} & \regstat{0}{110} & \regstat{0}{168} & \regstat{0}{108} & \regstat{0}{167}\\

\midrule\end{tabular}
\endgroup

\par\nointerlineskip

\begingroup
\linespread{1.0}\fontsize{8.5}{9.8}\selectfont
\setlength{\tabcolsep}{2.2pt}
\renewcommand{\arraystretch}{1.0}

\begin{tabular}{@{}L{0.37\linewidth}*{4}{C{\dimexpr(\linewidth-0.37\linewidth-8\tabcolsep)/4\relax}}@{}}
\multicolumn{5}{@{}p{\linewidth}@{}}{\bfseries Panel C. Joint controls for market exposure and premarket returns}\\[0.5pt]
Variable & Baseline model & With 09:25 cum. return & With ES & With ES and 09:25 cum. return\\
\midrule
\rowlabel{Weekend perpetual return} & \regpair{\regcoef{0}{536}{***}}{\regse{0}{193}} & \regpair{\regcoef{-0}{028}{}}{\regse{0}{018}} & \regpair{\regcoef{0}{338}{**}}{\regse{0}{140}} & \regpair{\regcoef{-0}{019}{}}{\regse{0}{017}}\\
\rowlabel{09:25 cumulative return} &  & \regpair{\regcoef{0}{993}{***}}{\regse{0}{016}} &  & \regpair{\regcoef{0}{999}{***}}{\regse{0}{016}}\\
\rowlabel{ES market exposure} &  &  & \regpair{\regcoef{0}{852}{**}}{\regse{0}{367}} & \regpair{\regcoef{-0}{054}{}}{\regse{0}{046}}\\

\addlinespace[0.5pt]
\rowlabel{Stock-weeks} & 2,848 & 2,848 & 2,848 & 2,848\\
\rowlabel{Stocks} & 267 & 267 & 267 & 267\\
\rowlabel{Perpetual incr. within-\ensuremath{R^2} (\%)} & \regstat{4}{931} & \regstat{0}{013} & \regstat{1}{798} & \regstat{0}{005}\\

\bottomrule\end{tabular}
\endgroup

\begin{exhibitnotes}
All models include historical stock characteristics, stock fixed effects, and reopening-date fixed effects. Panel A uses cumulative stock returns at the regular-session open and five minutes after the overnight open as dependent variables. Each pair of specifications compares estimates with and without ES market exposure $M_{iw}$ in a common sample. Panels B and C use cumulative opening returns. Panel B conditions on the full weekend ES return for an ex post comparison. In Panel C, controlling jointly for ES market exposure and the 09:25 cumulative stock return yields a remaining opening effect of $-1.86$ basis points per one-percentage-point higher weekend perpetual return, with a 90\% confidence interval of [$-4.69$, 0.97] basis points. The effect passes the equivalence test with bounds of $\pm5$ basis points.
\incrnote{}\regnote{}
\end{exhibitnotes}
\end{table}

\subsection{Robustness Tests}

The positive relation between weekend perpetual returns and cumulative opening returns is generally stable across alternative specifications, return measures, and sample restrictions. The perpetual return coefficient remains significantly positive with industry-by-reopening-date fixed effects, with the historical equal-weighted stock market beta and its interaction with Bitcoin (BTC) returns, and with alternative return starting points, price measures, and treatments of extreme observations (\apptabref{tab:D2}).

The wild cluster bootstrap at the calendar-week level supports the statistical significance of the baseline coefficient, and estimates obtained by excluding one reopening date at a time support its stability. Within-industry permutation tests show that the coefficient under actual security matching exceeds the coefficients obtained under random matching (\apptabref{tab:D3} and \appfigref{fig:D1}). The perpetual return coefficient is significantly positive at most individual venues (\apptabref{tab:D4}).

\FloatBarrier
\section{Conclusion}\label{sec:conclusion}

Using transaction data on equity perpetuals covering 421 U.S.-listed stocks across 22 venues, we examine the informational role of round-the-clock equity derivatives during stock market closures and reopening. Weekend perpetual returns are significantly positively associated with subsequent stock returns. Perpetual prices formed before stock index futures reopen already contain information about cumulative opening returns. In the baseline sample, a one-percentage-point higher weekend perpetual return is associated with a 59.2-basis-point higher cumulative stock return at the regular-session open, on average. Estimates from the expanded stock sample and results obtained by excluding each trading venue in turn further support this positive relationship.

Stock price adjustments associated with weekend perpetual returns are already apparent early in the overnight session. The positive relation between weekend perpetual returns and cumulative stock returns persists through the close of the trading day following the reopening day. Trade- and quote-level evidence further shows that within-minute price updates after the first transaction contain information about cumulative opening returns. Conditioning on cumulative stock returns early in the overnight session substantially reduces the incremental explanatory power of weekend perpetual returns, although the remaining opening effect at that stage is estimated imprecisely. In the sample with stock prices available five minutes before the regular-session open, conditioning on the cumulative stock return at that time yields a remaining opening effect that passes an equivalence test with bounds of $\pm10$ basis points per percentage point of perpetual return.

These findings show that prices formed in equity perpetual markets during stock market closures contain information about subsequent stock prices. Their incremental value depends on the stock price information available after trading resumes. By expanding coverage across stocks and trading venues and combining overnight, premarket, and regular-session price records, our analysis provides evidence on the link between weekend price information and early stock price discovery.

\clearpage
\appendix
\titleformat{\section}{\appendixsectionfont\raggedright}{Appendix~\thesection}{1em}{}
\titleformat{\subsection}{\appendixsubsectionfont\raggedright}{\thesubsection}{.5em}{}
\renewcommand{\thesection}{\Alph{section}}
\renewcommand{\thesubsection}{\thesection\arabic{subsection}}
\renewcommand{\tablename}{Appendix Table}
\renewcommand{\figurename}{Appendix Figure}
\renewcommand{\thetable}{\thesection\arabic{table}}
\renewcommand{\thefigure}{\thesection\arabic{figure}}
\renewcommand{\theHtable}{appendix.\thesection.\arabic{table}}
\renewcommand{\theHfigure}{appendix.\thesection.\arabic{figure}}
\section{Data Sources, Sample Construction, and Variable Definitions}\label{app:A}
\setcounter{table}{0}\setcounter{figure}{0}

This appendix documents the sample and data construction. \appref{app:A1} describes sample selection and coverage, and \appref{app:A2} lists the data sources. \appref{app:A3} defines return variables and historical stock characteristics, while \appref{app:A4} defines market exposure and trading characteristics. \appref{app:A5} describes price measurement and data processing.

\subsection{Sample Construction and Coverage}\label{app:A1}

We identify equity perpetuals using the contract directory as of November 19, 2026, at 10:35 New York time and historical listing records, and match them to the Nasdaq Trader securities directory for the same date. The sample is restricted to U.S.-listed common stocks and eligible ADRs. Listing announcements and verifiable historical trade records establish when each contract was available for trading. Inclusion requires that the contract and stock refer to the same security, have overlapping trading periods, and have valid transaction prices at the return endpoints. For trading suspensions or changes in security identity, we retain only observations whose security identity and trading period can be verified. The directory includes listed contracts and delisted contracts identified from verifiable listing and trading records. Coverage of previously delisted contracts depends on the availability of these records.

Trading channels that share the same underlying market are treated as a single venue, and duplicate market data records are removed.\footnote{Extended's order book and request-for-quote markets are treated as a single venue. The same rule applies to Vest Classic and Vest Markets.} The expanded transaction sample excludes Vest and Hyperliquid. The November 23 sample also uses the November 19 contract directory.

\apptabref{tab:A1} reports the size and coverage of the research samples.

\par\medskip\noindent\begin{minipage}{\linewidth}
\centering\setlength{\parskip}{0pt}
\captionof{table}{Sample construction and data coverage}\label{tab:A1}

\begingroup
\linespread{1.0}\fontsize{9.5}{11}\selectfont
\setlength{\tabcolsep}{2.2pt}
\renewcommand{\arraystretch}{1.0}

\begin{tabular}{@{}L{0.45\linewidth}*{4}{C{\dimexpr(\linewidth-0.45\linewidth-8\tabcolsep)/4\relax}}@{}}
\toprule
\multicolumn{5}{@{}p{\linewidth}@{}}{\bfseries Panel A. Research samples}\\[2pt]
Sample & Stocks & Venues & Stock-weeks & \grouphead{Stock-week\\coverage}\\
\midrule
\rowlabel{Expanded sample} & 421 & 22 & 7,489 & 100.0\%\\
\rowlabel{Common return starting point} & 319 & 22 & 4,079 & 54.5\%\\
\rowlabel{Baseline sample} & 319 & 22 & 4,060 & 54.2\%\\
\rowlabel{Valid 09:25 prices} & 284 & 22 & 3,708 & 49.5\%\\
\rowlabel{Valid overnight +5 min prices} & 225 & 22 & 2,859 & 38.2\%\\
\rowlabel{Valid 04:05 prices} & 299 & 22 & 3,945 & 52.7\%\\
\rowlabel{Common daytime sample} & 182 & 22 & 2,250 & 30.0\%\\
\rowlabel{Common overnight and daytime sample} & 118 & 22 & 1,408 & 18.8\%\\
\midrule
\end{tabular}
\endgroup

\par\nointerlineskip

\begingroup
\linespread{1.0}\fontsize{9.5}{11}\selectfont
\setlength{\tabcolsep}{2.2pt}
\renewcommand{\arraystretch}{1.0}

\begin{tabular}{@{}L{0.45\linewidth}*{2}{C{\dimexpr(\linewidth-0.45\linewidth-4\tabcolsep)/2\relax}}@{}}
\multicolumn{3}{@{}p{\linewidth}@{}}{\bfseries Panel B. ES market exposure sample}\\[2pt]
Sample & Stocks & Stock-weeks\\
\midrule
\rowlabel{Baseline sample} & 319 & 4,060\\
\rowlabel{ES returns available} & 293 & 3,096\\
\rowlabel{ES returns and historical SPY betas available} & 293 & 3,091\\
\bottomrule
\end{tabular}
\endgroup

\begin{exhibitnotes}
Panel A reports sample sizes, with stock-week coverage measured relative to the expanded sample of 7,489 stock-weeks. The sample with a common return starting point requires both stock and perpetual prices at Friday 20:00; the baseline sample also requires historical stock characteristics. The common overnight and daytime sample covers overnight, premarket, and regular trading, whereas the common daytime sample covers premarket and regular trading. Each sample requires the relevant prices and controls. Panel B matches the baseline sample to weekend ES returns and historical SPY market betas. ES observations require valid futures prices and must fall outside contract rollover windows.
\end{exhibitnotes}
\end{minipage}\par\medskip

\par\medskip\noindent\begin{minipage}{\linewidth}
\subsection{Data Sources}\label{app:A2}
\apptabref{tab:A2} reports the original data frequency and sample coverage for each perpetual trading venue.\par\medskip
\centering\setlength{\parskip}{0pt}
\captionof{table}{Sources and coverage of equity perpetual transaction data}\label{tab:A2}
\begingroup
\linespread{1.0}\fontsize{9.5}{11}\selectfont
\setlength{\tabcolsep}{2.2pt}
\renewcommand{\arraystretch}{1.0}
\begin{tabular}{
@{}
L{\dimexpr 0.20\linewidth-1.60\tabcolsep\relax}
C{\dimexpr 0.16\linewidth-1.28\tabcolsep\relax}
C{\dimexpr 0.26\linewidth-2.08\tabcolsep\relax}
C{\dimexpr 0.14\linewidth-1.12\tabcolsep\relax}
C{\dimexpr 0.24\linewidth-1.92\tabcolsep\relax}
@{}
}
\toprule
Venue & \grouphead{Original\\frequency} & Coverage dates & Stocks & Venue--stock-weeks\\
\midrule
MEXC
& 5 min & 02/09--11/23 & 337 & 5,753\\
Bitget
& 1, 5 min & 02/09--11/23 & 252 & 4,377\\
Gate
& 5 min & 02/02--11/23 & 207 & 3,066\\
KuCoin
& 5 min & 03/16--11/16 & 129 & 2,173\\
Bitunix
& 5 min & 03/30--11/16 & 136 & 2,048\\
Bybit
& Trade-level & 04/27--11/16 & 172 & 1,733\\
BingX
& 5 min & 10/12--11/23 & 236 & 1,558\\
Phemex
& 5 min & 02/02--11/16 & 95 & 1,437\\
HTX
& 5 min & 05/18--11/16 & 118 & 1,271\\
XT
& 5 min & 03/30--11/16 & 110 & 1,154\\
Binance
& 5 min & 03/30--11/23 & 154 & 1,009\\
Ourbit
& 5 min & 02/17--11/16 & 68 & 965\\
OKX
& 1, 5 min & 05/26--11/23 & 76 & 663\\
Deepcoin
& 5 min & 08/24--11/16 & 45 & 435\\
Helix
& 5 min & 05/26--11/16 & 97 & 369\\
BloFin
& 5 min & 06/08--11/16 & 28 & 227\\
edgeX
& 5 min & 07/27--11/16 & 19 & 216\\
Extended
& 5 min & 02/17--11/16 & 20 & 209\\
Deribit
& 5 min & 09/21--11/16 & 26 & 112\\
Aster
& 5 min & 03/09--11/23 & 28 & 103\\
Crypto.com
& 5 min & 06/01--11/16 & 26 & 51\\
Kraken
& 5 min & 09/08--11/16 & 6 & 23\\
\bottomrule
\end{tabular}
\endgroup

\begin{exhibitnotes}
The table reports observations in the expanded sample with valid weekend perpetual returns and corresponding stock prices, by venue. Coverage dates are the first and last valid reopening dates in 2026, measured in New York local time; coverage between these dates may be discontinuous. The 22 venues provide 28,952 venue--stock-weeks, which aggregate to 421 stocks and 7,489 stock-weeks. All venue trade data are standardized to five-minute bars.
\end{exhibitnotes}
\end{minipage}\par\medskip

Daily stock prices and minute-level premarket and after-hours bars come from Alpaca SIP (consolidated Securities Information Processor data), while overnight minute bars come from Alpaca BOATS. Individual trades and quotes come from these two feeds. The daily data used to construct historical stock characteristics begin in December 2025. ES futures data come from Yahoo Finance and Syncster. Daily SPY data and five-minute BTC spot data come from Yahoo Finance and Binance, respectively.

Industry classifications use SIC codes from SEC EDGAR. Corporate action data come from Alpaca and Yahoo Finance.

\par\medskip\noindent\begin{minipage}{\linewidth}
\subsection{Return Variables and Historical Stock Characteristics}\label{app:A3}
\centering\setlength{\parskip}{0pt}
\captionof{table}{Definitions of main return variables and historical stock characteristics}\label{tab:A3}
\begingroup\linespread{1.0}\fontsize{8.5}{9.7}\selectfont\setlength{\tabcolsep}{2.0pt}\renewcommand{\arraystretch}{1.0}
\begin{tabular}{@{}>{\raggedright\arraybackslash}p{\dimexpr 0.23000\linewidth-1.38000\tabcolsep\relax}>{\raggedright\arraybackslash}p{\dimexpr 0.47000\linewidth-2.82000\tabcolsep\relax}>{\raggedright\arraybackslash}p{\dimexpr 0.22000\linewidth-1.32000\tabcolsep\relax}>{\centering\arraybackslash}p{\dimexpr 0.08000\linewidth-0.48000\tabcolsep\relax}@{}}\toprule
\multicolumn{4}{@{}p{\linewidth}@{}}{\bfseries Panel A. Return variables and measures of incremental explanatory power}\\[0.5pt]
\multicolumn{1}{@{}L{\dimexpr 0.23000\linewidth-1.38000\tabcolsep\relax}}{Variable} & \multicolumn{1}{L{\dimexpr 0.47000\linewidth-2.82000\tabcolsep\relax}}{Definition} & \multicolumn{1}{L{\dimexpr 0.22000\linewidth-1.32000\tabcolsep\relax}}{Window and sample requirements} & \multicolumn{1}{C{\dimexpr 0.08000\linewidth-0.48000\tabcolsep\relax}@{}}{Unit}\\\midrule
Weekend perpetual return $W$ & Equal-weighted average of weekend log returns across trading venues with valid prices at both endpoints. & Friday 20:00 to Sunday 19:55 & \%\\
Pre-ES perpetual return & Equal-weighted venue returns from Friday 20:00 to Sunday 18:00, using venues that meet the full-period price requirements, venues selected as of 18:00, or historically active venues. & Through Sunday 18:00; historically active venues require valid prices at both endpoints on at least two of the preceding four calendar weekends & \%\\
Perpetual return over a weekend interval $W^{(k)}$ & Equal-weighted average of venue-level log returns over four nonoverlapping intervals; the same venues are used across intervals within each stock-week. & Four intervals from Friday 20:00 to Sunday 19:55 & \%\\
Cumulative stock return $R(t)$ & $100\ln[P^S(t)/P^S(t_0)]$; $t_0$ is Friday 20:00 in the baseline sample and 16:00 in the expanded sample. & Overnight, premarket, regular-session open, and subsequent times & \%\\
Cumulative return at the first overnight trade & Log return from the common starting point, measured using the opening price of the first BOATS minute with valid trades. & First hour after the overnight open on the evening before the actual reopening date & \%\\
Cumulative return at overnight +5 min & Log return from the common starting point, measured using the latest valid minute closing price within the first five minutes of overnight trading. & Five minutes from 20:00 on the evening before the actual reopening date & \%\\
Stock interval return & Stock log return over the specified trading interval. & Each specified trading interval & \%\\
Within-minute stock return & $100\ln(\text{closing price}/\text{opening price})$ for the first BOATS minute with valid trades. & First minute with valid trades within the first hour after the overnight open & \%\\
Incremental within-$R^2$ & Difference between the within-$R^2$ of the full and restricted models, multiplied by 100. & Same dependent variable, sample, fixed effects, and other regressors & \%\\
Partial $R^2$ & Difference in residual sums of squares between the restricted and full models, divided by the restricted model's residual sum of squares and multiplied by 100. & Same dependent variable, sample, fixed effects, and other regressors & \%\\
\midrule\end{tabular}\endgroup
\par\nointerlineskip
\begingroup\linespread{1.0}\fontsize{8.5}{9.7}\selectfont\setlength{\tabcolsep}{2.0pt}\renewcommand{\arraystretch}{1.0}
\begin{tabular}{@{}>{\raggedright\arraybackslash}p{\dimexpr 0.23000\linewidth-1.38000\tabcolsep\relax}>{\raggedright\arraybackslash}p{\dimexpr 0.47000\linewidth-2.82000\tabcolsep\relax}>{\raggedright\arraybackslash}p{\dimexpr 0.22000\linewidth-1.32000\tabcolsep\relax}>{\centering\arraybackslash}p{\dimexpr 0.08000\linewidth-0.48000\tabcolsep\relax}@{}}
\multicolumn{4}{@{}p{\linewidth}@{}}{\bfseries Panel B. Historical stock characteristics}\\[0.5pt]
\multicolumn{1}{@{}L{\dimexpr 0.23000\linewidth-1.38000\tabcolsep\relax}}{Variable} & \multicolumn{1}{L{\dimexpr 0.47000\linewidth-2.82000\tabcolsep\relax}}{Definition} & \multicolumn{1}{L{\dimexpr 0.22000\linewidth-1.32000\tabcolsep\relax}}{Window and sample requirements} & \multicolumn{1}{C{\dimexpr 0.08000\linewidth-0.48000\tabcolsep\relax}@{}}{Unit}\\\midrule
Friday regular-session return & $100\ln(C_{\rm Fri}/O_{\rm Fri})$, from the regular-session open to close. & Friday regular session of the same week & \%\\
Trailing 5-day return & Sum of $r_t$ over the last 5 trading days, requiring valid returns on all 5 days. & Through Friday, inclusive & \%\\
Trailing 20-day return & Sum of $r_t$ over the last 20 trading days, requiring valid returns on all 20 days. & Through Friday, inclusive & \%\\
Trailing 20-day volatility & Sample standard deviation of $r_t$, not annualized. & 20 trading days through Friday; at least 15 valid trading days & \%\\
Historical mean log dollar volume & Historical mean of daily $\ln(1+C_t V_t)$; $C_tV_t$ approximates daily dollar trading volume. & 20 trading days through Friday; at least 15 valid trading days & Log scale\\
\bottomrule\end{tabular}\endgroup
\begin{exhibitnotes}
Stock and perpetual returns are log price changes multiplied by 100. Incremental within-$R^2$ is measured in percentage points, and partial $R^2$ in percent. $C_t,O_t,V_t$ denote the unadjusted daily closing price, opening price, and share volume, respectively, and $r_t=100\ln(C_t/C_{t-1})$. Prices on adjacent trading days come from the same data source. Returns spanning identified stock split dates are set to missing.
\end{exhibitnotes}
\end{minipage}\par\medskip

\par\medskip\noindent\begin{minipage}{\linewidth}
\subsection{Market Exposure and Trading Characteristics}\label{app:A4}
\centering\setlength{\parskip}{0pt}
\captionof{table}{Definitions of market exposure and trading characteristics}\label{tab:A4}
\begingroup\linespread{1.0}\fontsize{8.5}{9.7}\selectfont\setlength{\tabcolsep}{2.0pt}\renewcommand{\arraystretch}{1.0}
\begin{tabular}{@{}>{\raggedright\arraybackslash}p{\dimexpr 0.23000\linewidth-1.38000\tabcolsep\relax}>{\raggedright\arraybackslash}p{\dimexpr 0.47000\linewidth-2.82000\tabcolsep\relax}>{\raggedright\arraybackslash}p{\dimexpr 0.22000\linewidth-1.32000\tabcolsep\relax}>{\centering\arraybackslash}p{\dimexpr 0.08000\linewidth-0.48000\tabcolsep\relax}@{}}\toprule
\multicolumn{4}{@{}p{\linewidth}@{}}{\bfseries Panel A. Market returns and stock-level exposure}\\[0.5pt]
\multicolumn{1}{@{}L{\dimexpr 0.23000\linewidth-1.38000\tabcolsep\relax}}{Variable} & \multicolumn{1}{L{\dimexpr 0.47000\linewidth-2.82000\tabcolsep\relax}}{Definition} & \multicolumn{1}{L{\dimexpr 0.22000\linewidth-1.32000\tabcolsep\relax}}{Window and sample requirements} & \multicolumn{1}{C{\dimexpr 0.08000\linewidth-0.48000\tabcolsep\relax}@{}}{Unit}\\\midrule
Weekend ES return $R^{ES}$ & $100\ln(F_{\rm Sun,19:55}/F_{\rm Fri,17:00})$, using closing prices of minute bars with positive volume ending exactly at the specified endpoints. & Friday 17:00 to Sunday 19:55 & \%\\
Historical SPY market beta & Market beta estimated from unadjusted stock close-to-close log returns and SPY close-to-close log returns, matched on common valid trading days and excluding known split dates. & 60 trading days through Friday; at least 30 common valid trading days & Unitless\\
ES market exposure $M$ & $\widehat\beta^{SPY} R^{ES}$. & Beta through Friday; ES through Sunday 19:55 & \%\\
Weekend BTC return $R^{BTC}$ & Log ratio of closing prices of the selected Friday and Sunday five-minute BTC/USDT bars, multiplied by 100. & Close of selected Friday and Sunday bars (see notes) & \%\\
Historical equal-weighted stock market beta & The market return is the equal-weighted average of available stock close-to-close log returns each day, after excluding observations with absolute daily log returns $\geq0.35$. A rolling beta is estimated for each stock using its returns and this market return. & 60 trading days through Friday; at least 30 observations & Unitless\\
Equal-weighted stock market beta$\times$BTC return & $\widehat\beta^M R^{BTC}$, included jointly with $\widehat\beta^M$ in the model with market beta and its interaction with the BTC return. & Current weekend (see notes) & \%\\
\midrule\end{tabular}\endgroup
\par\nointerlineskip
\begingroup\linespread{1.0}\fontsize{8.5}{9.7}\selectfont\setlength{\tabcolsep}{2.0pt}\renewcommand{\arraystretch}{1.0}
\begin{tabular}{@{}>{\raggedright\arraybackslash}p{\dimexpr 0.23000\linewidth-1.38000\tabcolsep\relax}>{\raggedright\arraybackslash}p{\dimexpr 0.47000\linewidth-2.82000\tabcolsep\relax}>{\raggedright\arraybackslash}p{\dimexpr 0.22000\linewidth-1.32000\tabcolsep\relax}>{\centering\arraybackslash}p{\dimexpr 0.08000\linewidth-0.48000\tabcolsep\relax}@{}}
\multicolumn{4}{@{}p{\linewidth}@{}}{\bfseries Panel B. Trading characteristics and cross-venue return dispersion}\\[0.5pt]
\multicolumn{1}{@{}L{\dimexpr 0.23000\linewidth-1.38000\tabcolsep\relax}}{Variable} & \multicolumn{1}{L{\dimexpr 0.47000\linewidth-2.82000\tabcolsep\relax}}{Definition} & \multicolumn{1}{L{\dimexpr 0.22000\linewidth-1.32000\tabcolsep\relax}}{Window and sample requirements} & \multicolumn{1}{C{\dimexpr 0.08000\linewidth-0.48000\tabcolsep\relax}@{}}{Unit}\\\midrule
Valid-return venues $B$ & Number of venues with valid transaction prices at both return endpoints for the same stock-week. & Endpoints of the full weekend return & Venue count\\
Cross-venue return dispersion $D$ & Sample standard deviation of weekend returns across at least two venues with valid returns. The interaction model subtracts the estimation-sample mean from dispersion, retaining its original units. & Current week, across venues & \%\\
Median absolute deviation (MAD) & Median absolute deviation of perpetual returns across venues, without rescaling for normality. & Current week, at least two venues & \%\\
Interquartile range (IQR) & Interquartile range of perpetual returns across venues. & Current week, at least two venues & \%\\
Historical perpetual trading activity $A$ & For each stock--venue pair, compute $L=\ln(1+\text{weekend volume})$. Subtract the mean of the six earlier observations from the most recent historical $L$ to measure volume relative to its historical level, then average equally across venues with valid observations. & Excludes the current week; at least 3 of the earlier 6 observations must be valid & Log difference\\
Historical overnight stock trading activity $E$ & Number of minutes with trades in the first 30 minutes after BOATS opens, divided by 30, then averaged across historical stock-weeks with valid data coverage. & Previous 6 stock-week observations; at least 3 valid; excludes the current week & Fraction\\
Sunday perpetual price observation lag $G$ & Time in minutes from the start of the latest valid five-minute bar to Sunday 19:55, averaged equally across venues. & Sunday 19:55 of the current week & Minutes\\
Nonlinear perpetual return terms & $|W|$, $W|W|$, in original units. & Full current weekend & \%; \ensuremath{(\%)^2}\\
\bottomrule\end{tabular}\endgroup
\begin{exhibitnotes}
In Panel A, the 0.35 screening threshold for the equal-weighted market return applies to daily log returns before multiplication by 100. The final five-minute BTC bar on Sunday starts at 19:55 and may close at 20:00. The model with market beta and its interaction with the BTC return therefore provides an ex post comparison that controls for common market information. In Panel B, volume is measured in each venue's original units. Weekend volume is the sum of volume in intervals with trades and requires data coverage for at least 95\% of the expected five-minute intervals. Historical trading activity uses the sequence of available stock-week or stock--venue-week observations; the historically active venue filter in \apptabref{tab:A3} uses four calendar weekends.
\end{exhibitnotes}
\end{minipage}\par\medskip

\subsection{Price Measurement and Data Processing}\label{app:A5}

\paragraph{Transaction prices and venue selection.}
We standardize perpetual trade data to five-minute bars. When the original data are one-minute bars or individual trades, the five-minute closing price is the last one-minute closing price or the last transaction price in the interval, respectively; volume is summed within the interval. Individual trades are assigned to intervals using UTC time. Valid bars require positive prices and volume. At cutoff $T$, the perpetual price is the most recent valid closing price among bars starting at $T-15$, $T-10$, and $T-5$ minutes. The stock price is the most recent valid minute closing price among bars starting at $T-5$ through $T-1$ minutes. The price observation lag is measured from the start of the bar to the cutoff.

The common set of venues used for cumulative and interval perpetual returns requires valid prices at Friday 20:00; Saturday 08:00 and 20:00; and Sunday 08:00, 12:00, 16:00, 18:00, 19:00, and 19:55. For perpetual returns before ES reopens, the two alternative venue selection rules---valid venues as of 18:00 and historically active venues---use trade records up to and including Sunday 18:00. Both samples end on November 16, and the contract universe is based on the November 19 directory. Stock price observation times follow the New York trading calendar and the actual reopening schedule.

\paragraph{Trade- and quote-level measures.}
The transaction-level sample is stratified by reopening date and whether the first trade occurs within five minutes of the overnight open. Each stratum contains at most 10 stock-weeks, for a total of 718. Within the first hour after the actual overnight open, we identify the earliest BOATS minute with valid opening and closing prices, positive trading volume, and a positive trade count. Clock-minute intervals include their left endpoint and exclude their right endpoint. Valid trades require positive prices and share quantities; duplicate records are removed using trade timestamps and identifiers. Valid two-sided quotes require a positive bid price and positive bid and ask sizes, with the ask price no lower than the bid.

We match the first and last trades within that minute to the most recent valid quote at the trade timestamp or within the preceding five seconds. The matched midpoint return is calculated only when both trades have valid quote matches. Within-minute VWAPs are calculated separately for the first and second 30-second intervals, with trades at the 30-second boundary assigned to the second interval. When both intervals contain valid trades, the within-minute stock return is the log ratio of the second VWAP to the first. Odd-lot trades are identified using the data provider's sale condition I. The alternative official opening price is taken from SIP trades on the primary listing exchange between 09:30 and 09:32 with sale conditions that include Q, requiring a unique positive qualifying price. This sample contains 263 stock-weeks for 9 stocks; a separate premarket two-sided quote sample covers 20 stocks.

\paragraph{Corporate actions.}
The baseline analysis calculates returns from unadjusted prices. Daily, overnight, and cumulative returns spanning identified stock splits or reverse splits are set to missing. Daily prices on Friday and the reopening date come from the same source. Alternative dividend treatments either add the cash dividend per share to the reopening-day opening price or exclude observations on the corresponding ex-dividend dates. Other corporate actions are identified from available event records.

\paragraph{Stock index futures and BTC returns.}
The baseline ES series uses five-minute trade bars for the ESZ26 contract from Yahoo Finance where available, supplemented with Syncster continuous futures minute bars for missing observations. The futures sample excludes rollover windows in March, June, and September when Friday falls between the 11th and 21st of the month. Trading arrangements and calendars are described in the CME contract specifications and trading calendar.

\apptabref{tab:A4} defines the ES and BTC return windows. The sample for the model with market beta and its interaction with the BTC return ends on November 16.

\paragraph{Return dispersion and average marginal effects.}
In \eqnref{eq:heterogeneity}, the marginal effect of the weekend perpetual return at dispersion $d$ is
\[
\beta+\lambda(d-\bar D)+\psi'Z_{iw}+a\operatorname{sgn}(W_{iw})+2b|W_{iw}|,\qquad W_{iw}\ne0.
\]
For each of the 2,109 stock-weeks with multiple venues, we evaluate the marginal effect at the 25th, 50th, 75th, and 90th percentiles of the dispersion distribution (0.0656, 0.1141, 0.1991, and 0.3718 percentage points), holding other variables at their observed values. We then average these marginal effects over the estimation sample. Standard errors and confidence intervals are calculated from the covariance matrix of the estimated coefficients.

\clearpage
\section{Supplementary Analysis of Weekend Perpetual Prices and Stock Market Reopenings}\label{app:B}
\setcounter{table}{0}\setcounter{figure}{0}

\par\medskip\noindent\begin{minipage}{\linewidth}
\subsection{Timing of Weekend Price Information}\label{app:B1}
\centering\setlength{\parskip}{0pt}
\captionof{table}{Timing of weekend perpetual returns and supplementary tests}\label{tab:B1}
\begingroup\linespread{1.0}\fontsize{8}{8.8}\selectfont\setlength{\tabcolsep}{2.2pt}\renewcommand{\arraystretch}{1.0}
\begin{tabular}{@{}L{0.36\linewidth}*{4}{C{\dimexpr(\linewidth-0.36\linewidth-8\tabcolsep)/4\relax}}@{}}\toprule
\multicolumn{5}{@{}p{\linewidth}@{}}{\bfseries Panel A. Perpetual returns before ES reopens under alternative venue selection rules}\\[0.5pt]
Variable & \grouphead{First overnight\\price} & \grouphead{Overnight\\+5 min} & 09:25 & \grouphead{Regular-session\\open}\\\midrule
\rowlabel{Venues with valid full-window returns} & \regpair{\regcoef{0}{432}{***}}{\regse{0}{067}} & \regpair{\regcoef{0}{577}{***}}{\regse{0}{064}} & \regpair{\regcoef{0}{360}{***}}{\regse{0}{129}} & \regpair{\regcoef{0}{318}{**}}{\regse{0}{124}}\\
\rowlabel{Venues with valid returns through 18:00} & \regpair{\regcoef{0}{437}{***}}{\regse{0}{067}} & \regpair{\regcoef{0}{585}{***}}{\regse{0}{065}} & \regpair{\regcoef{0}{359}{***}}{\regse{0}{128}} & \regpair{\regcoef{0}{317}{**}}{\regse{0}{123}}\\
\rowlabel{Historically active venues} & \regpair{\regcoef{0}{437}{***}}{\regse{0}{067}} & \regpair{\regcoef{0}{576}{***}}{\regse{0}{064}} & \regpair{\regcoef{0}{354}{***}}{\regse{0}{129}} & \regpair{\regcoef{0}{313}{**}}{\regse{0}{125}}\\
\addlinespace[0.5pt]
\rowlabel{Stock-weeks} & 3,069 & 2,446 & 3,152 & 3,436\\
\rowlabel{Stocks} & 244 & 201 & 253 & 286\\
\midrule\end{tabular}\endgroup
\par\nointerlineskip
\begingroup\linespread{1.0}\fontsize{8}{8.8}\selectfont\setlength{\tabcolsep}{2.2pt}\renewcommand{\arraystretch}{1.0}
\begin{tabular}{@{}L{0.40\linewidth}*{2}{C{\dimexpr(\linewidth-0.40\linewidth-4\tabcolsep)/2\relax}}@{}}
\multicolumn{3}{@{}p{\linewidth}@{}}{\bfseries Panel B. Industry-by-reopening-date fixed effects}\\[0.5pt]
Variable & \grouphead{Stock and reopening-date\\fixed effects} & \grouphead{Stock and industry$\times$date\\fixed effects}\\\midrule
\rowlabel{Fri. 20:00 to Sat. 20:00} & \regpair{\regcoef{0}{461}{***}}{\regse{0}{126}} & \regpair{\regcoef{0}{335}{***}}{\regse{0}{093}}\\
\rowlabel{Sat. 20:00 to Sun. 12:00} & \regpair{\regcoef{0}{497}{**}}{\regse{0}{217}} & \regpair{\regcoef{0}{365}{**}}{\regse{0}{158}}\\
\rowlabel{Sun. 12:00 to 18:00} & \regpair{\regcoef{0}{577}{**}}{\regse{0}{225}} & \regpair{\regcoef{0}{385}{**}}{\regse{0}{181}}\\
\rowlabel{Sun. 18:00 to 19:55} & \regpair{\regcoef{1}{039}{***}}{\regse{0}{277}} & \regpair{\regcoef{0}{746}{***}}{\regse{0}{208}}\\
\rowlabel{Final minus first coefficient} & \regpair{\regcoef{0}{578}{**}}{\regse{0}{217}} & \regpair{\regcoef{0}{411}{**}}{\regse{0}{164}}\\
\addlinespace[0.5pt]
\rowlabel{Stock-weeks} & 3,174 & 3,174\\
\rowlabel{Stocks} & 172 & 172\\
\rowlabel{Within-\ensuremath{R^2}} & \regstat{0}{100} & \regstat{0}{058}\\
\midrule\end{tabular}\endgroup
\par\nointerlineskip
\begingroup\linespread{1.0}\fontsize{8}{8.8}\selectfont\setlength{\tabcolsep}{1.6pt}\renewcommand{\arraystretch}{1.0}
\begin{tabular}{@{}L{0.34\linewidth}*{8}{C{\dimexpr(\linewidth-0.34\linewidth-16\tabcolsep)/8\relax}}@{}}
\multicolumn{9}{@{}p{\linewidth}@{}}{\bfseries Panel C. Cumulative perpetual returns at different weekend cutoffs}\\[0.5pt]
 & \multicolumn{2}{C{\dimexpr ((\linewidth-0.34\linewidth-16\tabcolsep)/8)*2+2\tabcolsep\relax}}{Saturday} & \multicolumn{6}{C{\dimexpr ((\linewidth-0.34\linewidth-16\tabcolsep)/8)*6+10\tabcolsep\relax}@{}}{Sunday}\\
\cmidrule(lr){2-3}\cmidrule(lr){4-9}
Variable & 08:00 & 20:00 & 08:00 & 12:00 & 16:00 & 18:00 & 19:00 & 19:55\\\midrule
\rowlabel{Cumulative perpetual return} & \regpair{\regcoef{-0}{023}{}}{\regse{0}{157}} & \regpair{\regcoef{0}{239}{**}}{\regse{0}{106}} & \regpair{\regcoef{0}{365}{**}}{\regse{0}{142}} & \regpair{\regcoef{0}{338}{**}}{\regse{0}{145}} & \regpair{\regcoef{0}{290}{*}}{\regse{0}{156}} & \regpair{\regcoef{0}{361}{**}}{\regse{0}{142}} & \regpair{\regcoef{0}{549}{***}}{\regse{0}{170}} & \regpair{\regcoef{0}{627}{***}}{\regse{0}{169}}\\
\addlinespace[0.5pt]
\rowlabel{Stock-weeks} & 3,838 & 3,838 & 3,838 & 3,838 & 3,838 & 3,838 & 3,838 & 3,838\\
\rowlabel{Within-\ensuremath{R^2}} & \regstat{0}{012} & \regstat{0}{021} & \regstat{0}{034} & \regstat{0}{032} & \regstat{0}{028} & \regstat{0}{037} & \regstat{0}{070} & \regstat{0}{092}\\
\rowlabel{Incr. within-\ensuremath{R^2} (\%)} & \regstat{0}{005} & \regstat{0}{830} & \regstat{2}{154} & \regstat{1}{919} & \regstat{1}{519} & \regstat{2}{441} & \regstat{5}{782} & \regstat{7}{919}\\
\midrule\end{tabular}\endgroup
\par\nointerlineskip
\begingroup\linespread{1.0}\fontsize{8}{8.8}\selectfont\setlength{\tabcolsep}{2.2pt}\renewcommand{\arraystretch}{1.0}
\begin{tabular}{@{}L{0.36\linewidth}*{3}{C{\dimexpr(\linewidth-0.36\linewidth-6\tabcolsep)/3\relax}}@{}}
\multicolumn{4}{@{}p{\linewidth}@{}}{\bfseries Panel D. Equal-length intervals before and after ES reopens}\\[0.5pt]
Variable & \grouphead{All valid observations\\No ES control} & \grouphead{Common ES sample\\No ES control} & \grouphead{Common ES sample\\ES control}\\\midrule
\rowlabel{Post- minus pre-ES coefficient} & \regpair{\regcoef{0}{338}{}}{\regse{0}{296}} & \regpair{\regcoef{0}{339}{}}{\regse{0}{307}} & \regpair{\regcoef{0}{257}{}}{\regse{0}{322}}\\
\addlinespace[0.5pt]
\rowlabel{Stock-weeks} & 3,900 & 2,968 & 2,968\\
\rowlabel{Stocks} & 311 & 288 & 288\\
\rowlabel{Within-\ensuremath{R^2}} & \regstat{0}{076} & \regstat{0}{089} & \regstat{0}{159}\\
\rowlabel{Incr. within-\ensuremath{R^2} (\%)} & \regstat{0}{227} & \regstat{0}{209} & \regstat{0}{120}\\
\bottomrule\end{tabular}\endgroup
\begin{exhibitnotes}\fontsize{8}{9.4}\selectfont 
Panel A reports coefficients on perpetual returns formed before ES reopens under three venue selection rules. For each stock return endpoint, all three rules use a common sample with valid prices under every rule. Panel B compares reopening-date and industry-by-reopening-date fixed effects in the same sample with industry classifications. Panel C estimates eight cumulative perpetual return coefficients using the same 3,838 stock-weeks and venues. Panel D jointly includes perpetual returns over three 115-minute intervals on Sunday (14:10--16:05, 16:05--18:00, and 18:00--19:55) and compares the coefficients for the last two intervals. Its final two columns use the same sample with ES data. Panels B, C, and D use cumulative opening returns as the dependent variable. All models control for historical stock characteristics and include stock and the corresponding date fixed effects. Incremental within-$R^2$ refers to cumulative perpetual returns in Panel C and the reparameterized coefficient-difference term in Panel D. \incrnote{}\regnote{}
\end{exhibitnotes}
\end{minipage}\par\medskip

\par\medskip\noindent\begin{minipage}{\linewidth}
\subsection{Additional Price Observation Times after Reopening}\label{app:B2}
\centering\setlength{\parskip}{0pt}
\captionof{table}{Cumulative stock returns after reopening: additional observation times}\label{tab:B2}
\begingroup\linespread{1.0}\fontsize{9.5}{11}\selectfont\setlength{\tabcolsep}{2.0pt}\renewcommand{\arraystretch}{1.0}
\begin{tabular}{@{}L{0.37\linewidth}*{5}{C{\dimexpr(\linewidth-0.37\linewidth-10\tabcolsep)/5\relax}}@{}}\toprule
\multicolumn{6}{@{}p{\linewidth}@{}}{\bfseries Cumulative stock returns in the common overnight and daytime sample}\\[2pt]
Variable & \grouphead{Overnight\\+60 min} & 06:00 & 09:00 & 09:15 & \grouphead{Regular open\\+30 min}\\\midrule
\rowlabel{Weekend perpetual return} & \regpair{\regcoef{0}{571}{***}}{\regse{0}{079}} & \regpair{\regcoef{0}{607}{***}}{\regse{0}{130}} & \regpair{\regcoef{0}{583}{***}}{\regse{0}{167}} & \regpair{\regcoef{0}{586}{***}}{\regse{0}{161}} & \regpair{\regcoef{0}{514}{***}}{\regse{0}{171}}\\
\addlinespace[2pt]
\rowlabel{Stock-weeks} & 1,408 & 1,408 & 1,408 & 1,408 & 1,408\\
\rowlabel{Stocks} & 118 & 118 & 118 & 118 & 118\\
\rowlabel{Within-\ensuremath{R^2}} & \regstat{0}{259} & \regstat{0}{144} & \regstat{0}{093} & \regstat{0}{089} & \regstat{0}{041}\\
\rowlabel{Perpetual incr. within-\ensuremath{R^2} (\%)} & \regstat{24}{480} & \regstat{13}{920} & \regstat{8}{742} & \regstat{8}{338} & \regstat{3}{132}\\
\bottomrule\end{tabular}\endgroup
\begin{exhibitnotes}
All columns use the common overnight and daytime sample of 1,408 stock-weeks in Panel A of \tabref{tab:5}. The dependent variable is the cumulative stock return from Friday 20:00 to the indicated observation time. All models control for historical stock characteristics and include stock and reopening-date fixed effects. \incrnote{}\regnote{}
\end{exhibitnotes}
\end{minipage}\par\medskip

\par\medskip\noindent\begin{minipage}{\linewidth}
\subsection{Incremental Information in Perpetual Returns and Equivalence Tests}\label{app:B3}
\centering\setlength{\parskip}{0pt}
\captionof{table}{Remaining opening effects, equivalence tests, and alternative price measures}\label{tab:B3}
\begingroup\linespread{1.0}\fontsize{9}{10.4}\selectfont\setlength{\tabcolsep}{2.2pt}\renewcommand{\arraystretch}{1.0}
\begin{tabular}{@{}L{\dimexpr 0.230\linewidth-2.300\tabcolsep\relax}C{\dimexpr 0.220\linewidth-2.200\tabcolsep\relax}C{\dimexpr 0.220\linewidth-2.200\tabcolsep\relax}C{\dimexpr 0.100\linewidth-1.000\tabcolsep\relax}C{\dimexpr 0.115\linewidth-1.150\tabcolsep\relax}C{\dimexpr 0.115\linewidth-1.150\tabcolsep\relax}@{}}\toprule
\multicolumn{6}{@{}p{\linewidth}@{}}{\bfseries Panel A. Equivalence tests of the remaining opening effect}\\[2pt]
 & \multicolumn{2}{c}{\grouphead{90\% confidence intervals (bp)}} & \multicolumn{3}{c}{\grouphead{Equivalence test $p$-values}}\\
\cmidrule(lr){2-3}\cmidrule(lr){4-6}
\grouphead{Stock return\\control time} & \grouphead{Two-way\\clustering} & \grouphead{Calendar-week\\wild cluster bootstrap} & \ensuremath{\pm}5 bp & \ensuremath{\pm}10 bp & \ensuremath{\pm}20 bp\\\midrule
\rowlabel{Overnight +5 min} & \tabnum{[-27.06, 13.35]} & \tabnum{[-30.11, 16.31]} & \tabnum{0.5611} & \tabnum{0.3973} & \tabnum{0.1399}\\
\rowlabel{04:05} & \tabnum{[-14.63, 11.06]} & \tabnum{[-17.23, 13.68]} & \tabnum{0.3376} & \tabnum{0.1438} & \tabnum{0.01089}\\
\rowlabel{09:25} & \tabnum{[-6.98, -0.40]} & \tabnum{[-7.47, 0.16]} & \tabnum{0.2533} & \tabnum{0.001251} & \tabnum{1.6\times10^{-10}}\\
\midrule\end{tabular}\endgroup
\par\nointerlineskip
\begingroup\linespread{1.0}\fontsize{9}{10.4}\selectfont\setlength{\tabcolsep}{2.2pt}\renewcommand{\arraystretch}{1.0}
\begin{tabular}{@{}L{0.37\linewidth}*{3}{C{\dimexpr(\linewidth-0.37\linewidth-6\tabcolsep)/3\relax}}@{}}
\multicolumn{4}{@{}p{\linewidth}@{}}{\bfseries Panel B. Remaining opening effects in alternative samples}\\[2pt]
Variable & \grouphead{Overnight +5 min\\All valid observations} & \grouphead{04:05\\Common daytime sample} & \grouphead{04:05\\All valid observations}\\\midrule
\rowlabel{Weekend perpetual return} & \regpair{\regcoef{0}{064}{}}{\regse{0}{145}} & \regpair{\regcoef{-0}{018}{}}{\regse{0}{076}} & \regpair{\regcoef{-0}{024}{}}{\regse{0}{086}}\\
\addlinespace[2pt]
\rowlabel{Stock-weeks} & 2,859 & 2,250 & 3,945\\
\rowlabel{Stocks} & 225 & 182 & 299\\
\rowlabel{Perpetual incr. within-\ensuremath{R^2} (\%)} & \regstat{0}{060} & \regstat{0}{005} & \regstat{0}{010}\\
\rowlabel{Perpetual partial \ensuremath{R^2} (\%)} & \regstat{0}{076} & \regstat{0}{010} & \regstat{0}{018}\\
\midrule\end{tabular}\endgroup
\par\nointerlineskip
\begingroup\linespread{1.0}\fontsize{9}{10.4}\selectfont\setlength{\tabcolsep}{2.2pt}\renewcommand{\arraystretch}{1.0}
\begin{tabular}{@{}L{0.37\linewidth}*{3}{C{\dimexpr(\linewidth-0.37\linewidth-6\tabcolsep)/3\relax}}@{}}
\multicolumn{4}{@{}p{\linewidth}@{}}{\bfseries Panel C. Alternative stock price measures}\\[2pt]
Dependent variable & \grouphead{Cumulative\\opening return} & \grouphead{09:25--09:35\\quote-midpoint return} & \grouphead{Cumulative return\\to 5-min VWAP}\\
Stock return control & \grouphead{09:25 cumulative\\quote-midpoint return} & \grouphead{09:25 cumulative\\quote-midpoint return} & \grouphead{09:25 cumulative\\stock return}\\\midrule
\rowlabel{Weekend perpetual return} & \regpair{\regcoef{-0}{056}{}}{\regse{0}{042}} & \regpair{\regcoef{-0}{246}{**}}{\regse{0}{086}} & \regpair{\regcoef{-0}{081}{**}}{\regse{0}{034}}\\
\addlinespace[2pt]
\rowlabel{Stock-weeks} & 365 & 365 & 3,708\\
\rowlabel{Stocks} & 20 & 20 & 284\\
\rowlabel{Within-\ensuremath{R^2}} & \regstat{0}{952} & \regstat{0}{114} & \regstat{0}{830}\\
\rowlabel{Perpetual incr. within-\ensuremath{R^2} (\%)} & \regstat{0}{034} & \regstat{2}{490} & \regstat{0}{096}\\
\bottomrule\end{tabular}\endgroup
\begin{exhibitnotes}
Panel A reports remaining opening effects in basis points per one-percentage-point higher weekend perpetual return. The three stock return control times correspond to 1,408, 2,250, and 3,708 stock-weeks, respectively. The first two columns report 90\% confidence intervals based on two-way clustering and the wild cluster bootstrap at the calendar-week level. The last three columns report two one-sided test $p$-values using two-way clustered standard errors. Panel B compares the coefficients on weekend perpetual returns in cumulative opening return regressions conditional on cumulative stock returns at the indicated times. In Panel C, columns (1) and (2) use, respectively, the return from the Friday 20:00 quote midpoint to the regular-session open and the 09:25--09:35 quote-midpoint return; both control for the cumulative quote-midpoint return at 09:25. Column (3) uses the cumulative stock return from Friday 20:00 to the 09:30--09:35 VWAP and controls for the cumulative stock return at 09:25. \incrnote{} Partial $R^2$ is reported in \%. \regnote{}
\end{exhibitnotes}
\end{minipage}\par\medskip

\par\medskip\noindent\begin{minipage}{\linewidth}
\subsection{Price Persistence and Intraday Changes in Weekend Return Coefficients}\label{app:B4}
\centering\setlength{\parskip}{0pt}
\captionof{table}{Persistence and intraday changes in stock price adjustments}\label{tab:B4}
\begingroup\linespread{1.0}\fontsize{9}{10.4}\selectfont\setlength{\tabcolsep}{2.2pt}\renewcommand{\arraystretch}{1.0}
\begin{tabular}{@{}L{0.38\linewidth}*{4}{C{\dimexpr(\linewidth-0.38\linewidth-8\tabcolsep)/4\relax}}@{}}\toprule
\multicolumn{5}{@{}p{\linewidth}@{}}{\bfseries Panel A. Longer-horizon cumulative returns and intraday coefficient differences}\\[2pt]
 & \multicolumn{2}{C{\dimexpr((\linewidth-0.38\linewidth-8\tabcolsep)/4)*2+2\tabcolsep\relax}}{Longer-horizon cumulative returns} & \multicolumn{2}{C{\dimexpr((\linewidth-0.38\linewidth-8\tabcolsep)/4)*2+2\tabcolsep\relax}@{}}{Open-to-close returns}\\
\cmidrule(lr){2-3}\cmidrule(lr){4-5}
Variable & \grouphead{(1)\\Day 3 close} & \grouphead{(2)\\Day 5 close} & (3) & (4)\\\midrule
\rowlabel{Weekend perpetual return} & \regpair{\regcoef{0}{336}{*}}{\regse{0}{198}} & \regpair{\regcoef{0}{208}{}}{\regse{0}{259}} & \textemdash{} & \textemdash{}\\
\rowlabel{Final minus first coefficient} & \textemdash{} & \textemdash{} & \regpair{\regcoef{-0}{491}{**}}{\regse{0}{199}} & \regpair{\regcoef{-0}{541}{**}}{\regse{0}{246}}\\
\addlinespace[2pt]
\rowlabel{ES market exposure in intraday regression} & \textemdash{} & \textemdash{} & NO & YES\\
\addlinespace[2pt]
\rowlabel{Stock-weeks} & 3,949 & 3,949 & 3,838 & 2,920\\
\rowlabel{Stocks} & 319 & 319 & 314 & 285\\
\rowlabel{Within-\ensuremath{R^2}} & \regstat{0}{016} & \regstat{0}{016} & \regstat{0}{016} & \regstat{0}{017}\\
\rowlabel{Perpetual incr. within-\ensuremath{R^2} (\%)} & \regstat{0}{197} & \regstat{0}{056} & \textemdash{} & \textemdash{}\\
\rowlabel{Difference term incr. within-\ensuremath{R^2} (\%)} & \textemdash{} & \textemdash{} & \regstat{0}{602} & \regstat{0}{664}\\
\midrule\end{tabular}\endgroup
\par\nointerlineskip
\begingroup\linespread{1.0}\fontsize{9}{10.4}\selectfont\setlength{\tabcolsep}{2.2pt}\renewcommand{\arraystretch}{1.0}
\begin{tabular}{@{}L{0.38\linewidth}*{3}{C{\dimexpr(\linewidth-0.38\linewidth-6\tabcolsep)/3\relax}}@{}}
\multicolumn{4}{@{}p{\linewidth}@{}}{\bfseries Panel B. Four perpetual return intervals and opening, closing, and intraday returns}\\[0.5pt]
Variable & \grouphead{Cumulative\\opening return} & \grouphead{Cumulative return\\to reopening day close} & \grouphead{Open-to-close\\return}\\\midrule
\rowlabel{Fri. 20:00 to Sat. 20:00} & \regpair{\regcoef{0}{480}{***}}{\regse{0}{120}} & \regpair{\regcoef{0}{678}{***}}{\regse{0}{132}} & \regpair{\regcoef{0}{198}{}}{\regse{0}{136}}\\
\rowlabel{Sat. 20:00 to Sun. 12:00} & \regpair{\regcoef{0}{502}{**}}{\regse{0}{200}} & \regpair{\regcoef{0}{528}{**}}{\regse{0}{234}} & \regpair{\regcoef{0}{026}{}}{\regse{0}{172}}\\
\rowlabel{Sun. 12:00 to 18:00} & \regpair{\regcoef{0}{606}{***}}{\regse{0}{216}} & \regpair{\regcoef{0}{907}{***}}{\regse{0}{228}} & \regpair{\regcoef{0}{300}{}}{\regse{0}{252}}\\
\rowlabel{Sun. 18:00 to 19:55} & \regpair{\regcoef{1}{132}{***}}{\regse{0}{268}} & \regpair{\regcoef{0}{839}{***}}{\regse{0}{294}} & \regpair{\regcoef{-0}{293}{}}{\regse{0}{180}}\\
\rowlabel{Final minus first coefficient} & \regpair{\regcoef{0}{652}{***}}{\regse{0}{235}} & \regpair{\regcoef{0}{161}{}}{\regse{0}{241}} & \regpair{\regcoef{-0}{491}{**}}{\regse{0}{199}}\\
\rowlabel{Coefficient difference $p$-value} & \regstat{0}{0083} & \regstat{0}{5075} & \regstat{0}{0182}\\
\addlinespace[0.5pt]
\rowlabel{Stock-weeks} & 3,838 & 3,838 & 3,838\\
\rowlabel{Stocks} & 314 & 314 & 314\\
\rowlabel{Within-\ensuremath{R^2}} & \regstat{0}{113} & \regstat{0}{050} & \regstat{0}{016}\\
\bottomrule\end{tabular}\endgroup
\begin{exhibitnotes}
Columns (1) and (2) of Panel A use cumulative stock returns from Friday 20:00 to the close of the third and fifth trading days after reopening as dependent variables and report the coefficients on weekend perpetual returns. The reopening day is excluded when counting subsequent trading days. Columns (3) and (4) use stock returns from the regular-session open to the close of the reopening day and report the difference between the final and first interval coefficients in regressions that jointly include all four weekend perpetual return intervals, without and with ES market exposure, respectively. Perpetual incremental within-$R^2$ refers to the full weekend return; incremental within-$R^2$ for the coefficient difference refers to the reparameterized difference term. Panel B estimates the joint four-interval regressions for cumulative opening returns, cumulative returns to the close of the reopening day, and open-to-close returns, using the same stock-weeks and venues. \incrnote{}\regnote{}
\end{exhibitnotes}
\end{minipage}\par\medskip
\clearpage
\section{Supplementary Evidence on Price Updates and Heterogeneity}\label{app:C}
\setcounter{table}{0}\setcounter{figure}{0}

\par\medskip\noindent\begin{minipage}{\linewidth}
\subsection{Alternative Price Measures for Within-Minute Returns}\label{app:C1}
\centering\setlength{\parskip}{0pt}
\captionof{table}{Within-minute stock returns with and without ES market exposure}\label{tab:C1}
\begingroup\linespread{1.0}\fontsize{9}{10.5}\selectfont\setlength{\tabcolsep}{2.2pt}\renewcommand{\arraystretch}{1.0}
\begin{tabular}{@{}L{0.35\linewidth}*{4}{C{\dimexpr(\linewidth-0.35\linewidth-8\tabcolsep)/4\relax}}@{}}\toprule
Variable & Within-minute quote midpoint & Quote midpoint matched to trades & 30-second VWAP & Excluding odd-lot trades\\\midrule
\rowlabel{Without ES market exposure} & \regpair{\regcoef{0}{182}{}}{\regse{0}{120}} & \regpair{\regcoef{0}{189}{}}{\regse{0}{161}} & \regpair{\regcoef{0}{073}{*}}{\regse{0}{040}} & \regpair{\regcoef{0}{144}{**}}{\regse{0}{056}}\\
\rowlabel{With ES market exposure} & \regpair{\regcoef{0}{140}{}}{\regse{0}{144}} & \regpair{\regcoef{0}{157}{}}{\regse{0}{198}} & \regpair{\regcoef{0}{051}{}}{\regse{0}{042}} & \regpair{\regcoef{0}{148}{**}}{\regse{0}{066}}\\
\addlinespace[2pt]
\rowlabel{Stock-weeks} & 567 & 394 & 358 & 571\\
\rowlabel{Stocks} & 189 & 166 & 129 & 191\\
\bottomrule\end{tabular}\endgroup
\begin{exhibitnotes}
This table reports the coefficient on weekend perpetual returns in regressions of within-minute stock returns. Each column compares specifications with and without ES market exposure in the common ES sample for the indicated price measure.\regnote{}
\end{exhibitnotes}
\end{minipage}\par\medskip

\par\medskip\noindent\begin{minipage}{\linewidth}
\subsection{Robustness Tests for Cross-Venue Return Dispersion}\label{app:C2}
\centering\setlength{\parskip}{0pt}
\captionof{table}{Cross-venue return dispersion and early price responses: robustness tests}\label{tab:C2}
\begingroup\linespread{1.0}\fontsize{8.5}{10.0}\selectfont\setlength{\tabcolsep}{1.6pt}\renewcommand{\arraystretch}{1.0}
\begin{tabular}{@{}L{0.28\linewidth}*{7}{C{\dimexpr(\linewidth-0.28\linewidth-14\tabcolsep)/7\relax}}@{}}\toprule
Variable & \grouphead{Baseline\\sample} & \grouphead{Common\\overnight\\and daytime\\sample} & \grouphead{Common\\daytime\\sample} & MAD & IQR & \grouphead{5-minute\\price\\window} & \grouphead{Exclude\\extreme\\1\%}\\\midrule
\rowlabel{Perpetual return \ensuremath{\times} dispersion} & \regpair{\regcoef{0}{215}{***}}{\regse{0}{048}} & \regpair{\regcoef{0}{220}{***}}{\regse{0}{044}} & \regpair{\regcoef{0}{219}{***}}{\regse{0}{051}} & \regpair{\regcoef{0}{283}{**}}{\regse{0}{113}} & \regpair{\regcoef{0}{221}{***}}{\regse{0}{055}} & \regpair{\regcoef{0}{247}{***}}{\regse{0}{060}} & \regpair{\regcoef{0}{203}{**}}{\regse{0}{100}}\\
\addlinespace[2pt]
\rowlabel{Stock-weeks} & 2,109 & 1,104 & 1,580 & 2,109 & 2,109 & 2,057 & 2,087\\
\rowlabel{Stocks} & 168 & 98 & 134 & 168 & 168 & 162 & 168\\
\rowlabel{Interaction incr. within-\ensuremath{R^2} (\%)} & \regstat{1}{367} & \regstat{1}{798} & \regstat{1}{628} & \regstat{1}{000} & \regstat{1}{176} & \regstat{1}{319} & \regstat{0}{602}\\
\bottomrule\end{tabular}\endgroup
\begin{exhibitnotes}
All columns use the cumulative stock return five minutes after the overnight open as the dependent variable and report the coefficient on the interaction between perpetual returns and the indicated dispersion measure in \eqnref{eq:heterogeneity}. The first three columns measure dispersion using the contemporaneous standard deviation of returns across venues and use, respectively, the baseline interaction sample, the common overnight and daytime sample, and the common daytime sample. The remaining columns use the unscaled median absolute deviation (MAD), the interquartile range (IQR), dispersion calculated with price observation lags of at most five minutes at both endpoints, and a sample excluding the extreme 1\% of return observations. The incremental within-$R^2$ for the interaction is the change in within-$R^2$ when the interaction is added.\incrnote{}\regnote{}
\end{exhibitnotes}
\end{minipage}\par\medskip
\clearpage
\section{Market Information, Robustness Tests, and Supplementary Analysis}\label{app:D}
\setcounter{table}{0}\setcounter{figure}{0}

This appendix presents equivalence tests that control jointly for ES market exposure and premarket returns, followed by robustness tests of the baseline results across model specifications, inference methods, and venue coverage.

\par\medskip\noindent\begin{minipage}{\linewidth}
\subsection{Equivalence Tests Controlling for Market Exposure}\label{app:D1}
\centering\setlength{\parskip}{0pt}
\captionof{table}{Equivalence tests with joint controls for ES market exposure and premarket returns}\label{tab:D1}
\begingroup\linespread{1.0}\fontsize{9}{10.5}\selectfont\setlength{\tabcolsep}{2.2pt}\renewcommand{\arraystretch}{1.0}
\begin{tabular}{@{}L{\dimexpr 0.280\linewidth-2.800\tabcolsep\relax}C{\dimexpr 0.160\linewidth-1.600\tabcolsep\relax}C{\dimexpr 0.160\linewidth-1.600\tabcolsep\relax}C{\dimexpr 0.100\linewidth-1.000\tabcolsep\relax}C{\dimexpr 0.145\linewidth-1.450\tabcolsep\relax}C{\dimexpr 0.155\linewidth-1.550\tabcolsep\relax}@{}}\toprule
 & \multicolumn{2}{c}{\grouphead{90\% confidence interval (bp)}} & \multicolumn{3}{c}{\grouphead{Equivalence-test $p$-value}}\\
\cmidrule(lr){2-3}\cmidrule(lr){4-6}
Controls & \grouphead{Two-way\\clustered} & \grouphead{Calendar-week\\wild bootstrap} & \ensuremath{\pm}5 bp & \ensuremath{\pm}10 bp & \ensuremath{\pm}20 bp\\\midrule
\rowlabel{09:25 cumulative return and ES market exposure} & \tabnum{[-4.69, 0.97]} & \tabnum{[-3.78, 0.01]} & \tabnum{0.03467} & \tabnum{1.534\times10^{-5}} & \tabnum{2.136\times10^{-12}}\\
\bottomrule\end{tabular}\endgroup
\begin{exhibitnotes}
This table uses 2,848 stock-weeks with both ES market exposure and the 09:25 cumulative stock return. It reports 90\% confidence intervals for the remaining opening effect, measured in basis points per one-percentage-point higher weekend perpetual return, using two-way clustering by stock and calendar week and a wild cluster bootstrap at the calendar-week level. The final three columns report $p$-values from two one-sided tests (TOST) at the indicated equivalence bounds.
\end{exhibitnotes}
\end{minipage}\par\medskip

\par\medskip\noindent\begin{minipage}{\linewidth}
\subsection{Robustness Tests of the Baseline Results}\label{app:D2}

\apptabref{tab:D2} examines the sensitivity of the baseline results to common market information, extreme observations, return measures, and venue coverage restrictions. Weekend perpetual returns are positively associated with cumulative opening returns at the 1\% significance level in every alternative specification.

\centering\setlength{\parskip}{0pt}
\captionof{table}{Robustness tests of the baseline results}\label{tab:D2}

\begingroup
\linespread{1.0}\fontsize{9}{10.4}\selectfont
\setlength{\tabcolsep}{2.0pt}
\renewcommand{\arraystretch}{1.0}

\begin{tabular}{@{}L{0.37\linewidth}*{6}{C{\dimexpr(\linewidth-0.37\linewidth-12\tabcolsep)/6\relax}}@{}}
\toprule
\multicolumn{7}{@{}p{\linewidth}@{}}{\bfseries Panel A. Common market information and extreme observations}\\[0.5pt]
Variable & \grouphead{Industry\\$\times$ date\\fixed effects} & \grouphead{Market beta\\$\times$ BTC\\return} & \grouphead{Winsorize\\0.5\%\\each tail} & \grouphead{Winsorize\\1\%\\each tail} & \grouphead{Exclude\\extreme\\1\%} & \grouphead{Exclude\\five extreme\\dates}\\
\midrule
\rowlabel{Weekend perpetual return} & \regpair{\regcoef{0}{393}{***}}{\regse{0}{128}} & \regpair{\regcoef{0}{384}{***}}{\regse{0}{130}} & \regpair{\regcoef{0}{540}{***}}{\regse{0}{154}} & \regpair{\regcoef{0}{496}{***}}{\regse{0}{138}} & \regpair{\regcoef{0}{530}{***}}{\regse{0}{170}} & \regpair{\regcoef{0}{404}{***}}{\regse{0}{112}}\\

\addlinespace[0.5pt]
\rowlabel{Stock-weeks} & 3,256 & 3,946 & 4,060 & 4,060 & 4,019 & 3,453\\
\rowlabel{Stocks} & 172 & 319 & 319 & 319 & 319 & 317\\
\rowlabel{Within-\ensuremath{R^2}} & \regstat{0}{044} & \regstat{0}{130} & \regstat{0}{070} & \regstat{0}{059} & \regstat{0}{056} & \regstat{0}{041}\\
\rowlabel{Perpetual incr. within-\ensuremath{R^2} (\%)} & \regstat{3}{576} & \regstat{3}{445} & \regstat{6}{081} & \regstat{5}{190} & \regstat{4}{638} & \regstat{3}{636}\\

\midrule\end{tabular}
\endgroup

\par\nointerlineskip

\begingroup
\linespread{1.0}\fontsize{9}{10.4}\selectfont
\setlength{\tabcolsep}{2.0pt}
\renewcommand{\arraystretch}{1.0}

\begin{tabular}{@{}L{0.37\linewidth}*{6}{C{\dimexpr(\linewidth-0.37\linewidth-12\tabcolsep)/6\relax}}@{}}
\multicolumn{7}{@{}p{\linewidth}@{}}{\bfseries Panel B. Alternative return starting points and price measures}\\[0.5pt]
 & \multicolumn{3}{C{\dimexpr ((\linewidth-0.37\linewidth-12\tabcolsep)/6)*3+4\tabcolsep\relax}}{Stock/perpetual return starting time} & \multicolumn{3}{C{\dimexpr ((\linewidth-0.37\linewidth-12\tabcolsep)/6)*3+4\tabcolsep\relax}@{}}{Perpetual return measure}\\
\cmidrule(lr){2-4}
\cmidrule(lr){5-7}
Variable & 16:00/20:00 & 20:00/20:00 & 16:00/16:00 & \grouphead{Cross-venue\\median} & \grouphead{5-minute\\window} & \grouphead{15-minute\\VWAP}\\
\midrule
\rowlabel{Weekend perpetual return} & \regpair{\regcoef{0}{474}{***}}{\regse{0}{139}} & \regpair{\regcoef{0}{476}{***}}{\regse{0}{136}} & \regpair{\regcoef{0}{589}{***}}{\regse{0}{104}} & \regpair{\regcoef{0}{591}{***}}{\regse{0}{164}} & \regpair{\regcoef{0}{613}{***}}{\regse{0}{168}} & \regpair{\regcoef{0}{674}{***}}{\regse{0}{173}}\\

\addlinespace[0.5pt]
\rowlabel{Stock-weeks} & 3,944 & 3,944 & 3,944 & 4,060 & 3,869 & 1,019\\
\rowlabel{Stocks} & 318 & 318 & 318 & 319 & 310 & 65\\
\rowlabel{Within-\ensuremath{R^2}} & \regstat{0}{057} & \regstat{0}{063} & \regstat{0}{106} & \regstat{0}{081} & \regstat{0}{089} & \regstat{0}{116}\\
\rowlabel{Perpetual incr. within-\ensuremath{R^2} (\%)} & \regstat{5}{166} & \regstat{5}{515} & \regstat{10}{086} & \regstat{7}{064} & \regstat{7}{622} & \regstat{11}{645}\\

\midrule\end{tabular}
\endgroup

\par\nointerlineskip

\begingroup
\linespread{1.0}\fontsize{9}{10.4}\selectfont
\setlength{\tabcolsep}{2.2pt}
\renewcommand{\arraystretch}{1.0}

\begin{tabular}{@{}L{0.37\linewidth}*{4}{C{\dimexpr(\linewidth-0.37\linewidth-8\tabcolsep)/4\relax}}@{}}
\multicolumn{5}{@{}p{\linewidth}@{}}{\bfseries Panel C. Venue coverage and sample selection}\\[0.5pt]
 & \multicolumn{1}{C{\dimexpr ((\linewidth-0.37\linewidth-8\tabcolsep)/4)*1+0\tabcolsep\relax}}{Listed venues} & \multicolumn{3}{C{\dimexpr ((\linewidth-0.37\linewidth-8\tabcolsep)/4)*3+4\tabcolsep\relax}@{}}{Valid-return venues}\\
\cmidrule(lr){2-2}
\cmidrule(lr){3-5}
Variable & \ensuremath{\geq}2 & \ensuremath{\geq}2 & \ensuremath{\geq}3 & \ensuremath{\geq}5\\
\midrule
\rowlabel{Weekend perpetual return} & \regpair{\regcoef{0}{600}{***}}{\regse{0}{167}} & \regpair{\regcoef{0}{656}{***}}{\regse{0}{183}} & \regpair{\regcoef{0}{699}{***}}{\regse{0}{198}} & \regpair{\regcoef{0}{469}{***}}{\regse{0}{139}}\\

\addlinespace[0.5pt]
\rowlabel{Stock-weeks} & 4,035 & 3,195 & 2,770 & 1,962\\
\rowlabel{Stocks} & 313 & 229 & 203 & 164\\
\rowlabel{Within-\ensuremath{R^2}} & \regstat{0}{082} & \regstat{0}{105} & \regstat{0}{111} & \regstat{0}{056}\\
\rowlabel{Perpetual incr. within-\ensuremath{R^2} (\%)} & \regstat{7}{151} & \regstat{8}{528} & \regstat{9}{287} & \regstat{4}{870}\\

\bottomrule\end{tabular}
\endgroup

\begin{exhibitnotes}
All models use cumulative opening returns (\%) as the dependent variable and include historical stock characteristics and stock fixed effects. The first column of Panel A includes two-digit SIC industry-by-reopening-date fixed effects; all other specifications include reopening-date fixed effects. The BTC specification includes the historical equal-weighted stock market beta and its interaction with BTC returns, using full weekend market data for an ex post conditional comparison. Winsorization applies to both weekend perpetual returns and cumulative opening returns. The extreme 1\% restriction excludes observations with $|W_{iw}|$ above its 99th percentile. The five extreme reopening dates are those with the largest values of $\sum_i W_{iw}^2$. The first three columns of Panel B compare return starting points in the same 3,944 stock-weeks. The remaining columns measure perpetual returns using the cross-venue median, a five-minute price observation window, and 15-minute VWAPs. Panel C varies the minimum number of listed venues or venues with valid returns.\incrnote{}\regnote{}
\end{exhibitnotes}
\end{minipage}\par\medskip

\par\medskip\noindent\begin{minipage}{\linewidth}
\subsection{Statistical Inference and Exclusion of Individual Reopening Dates}\label{app:D3}
The wild cluster bootstrap at the calendar-week level supports the statistical significance of the baseline coefficient. Excluding one reopening date at a time yields coefficients between 0.477 and 0.651 (\apptabref{tab:D3}). In the within-industry permutation tests, the perpetual return coefficient under actual security matching exceeds all 999 coefficients obtained under random matching (\appfigref{fig:D1}).
\par\medskip
\centering\setlength{\parskip}{0pt}
\captionof{table}{Permutation tests, wild cluster bootstrap inference, and exclusion of individual reopening dates}\label{tab:D3}
\begingroup\linespread{1.0}\fontsize{9}{10.4}\selectfont\setlength{\tabcolsep}{2.2pt}\renewcommand{\arraystretch}{1.0}
\begin{tabular}{@{}L{0.40\linewidth}*{4}{C{\dimexpr(\linewidth-0.40\linewidth-8\tabcolsep)/4\relax}}@{}}\toprule
\multicolumn{5}{@{}p{\linewidth}@{}}{\bfseries Panel A. Within-industry permutation and wild cluster bootstrap tests}\\[2pt]
Model & \grouphead{Actual\\matching\\coefficient} & \grouphead{Median\\permutation\\coefficient} & \grouphead{Permutation\\$p$-value} & \grouphead{Wild cluster\\bootstrap\\$p$-value}\\\midrule
\rowlabel{Baseline regression} & \tabnum{0.592} & \tabnum{0.250} & \tabnum{0.001} & \tabnum{0.0002}\\
\rowlabel{Common sample with ES market exposure} & \tabnum{0.320} & \tabnum{0.052} & \tabnum{0.001} & \tabnum{0.023}\\
\midrule\end{tabular}\endgroup
\par\nointerlineskip
\begingroup\linespread{1.0}\fontsize{9}{10.4}\selectfont\setlength{\tabcolsep}{2.2pt}\renewcommand{\arraystretch}{1.0}
\begin{tabular}{@{}L{0.40\linewidth}*{3}{C{\dimexpr(\linewidth-0.40\linewidth-6\tabcolsep)/3\relax}}@{}}
\multicolumn{4}{@{}p{\linewidth}@{}}{\bfseries Panel B. Estimates excluding one reopening date at a time}\\[2pt]
Sample adjustment & Regressions & Minimum coefficient & Maximum coefficient\\\midrule
\rowlabel{Exclude one reopening date at a time} & 40 & \tabnum{0.477} & \tabnum{0.651}\\
\bottomrule\end{tabular}\endgroup
\begin{exhibitnotes}
Panel A randomly permutes security matches 999 times within industry and reopening date and uses 9,999 wild cluster bootstrap draws at the calendar-week level for inference. \appfigref{fig:D1} shows the distributions of permutation coefficients. Panel B reestimates the baseline regression after excluding one reopening date at a time.
\end{exhibitnotes}
\end{minipage}\par\medskip

\par\addvspace{7pt}\noindent\begin{minipage}{\linewidth}\centering\includegraphics[width=.94\linewidth]{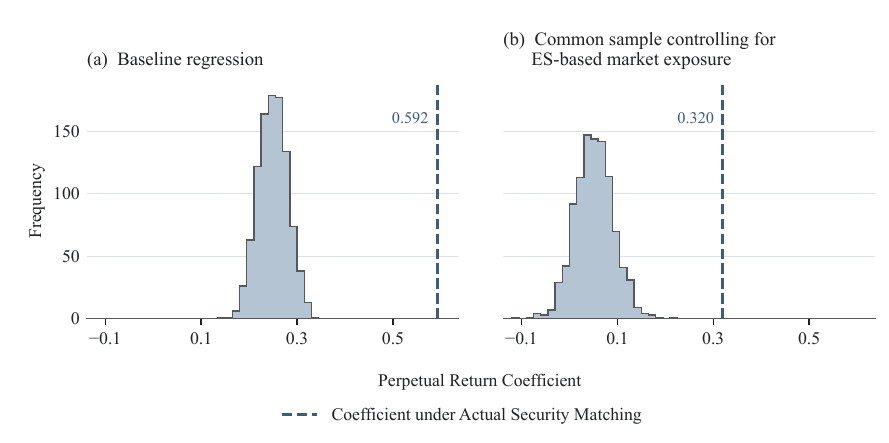}\captionof{figure}{Distribution of perpetual return coefficients under random security matching}\label{fig:D1}\begin{exhibitnotes}
This figure shows 999 perpetual return coefficients obtained by randomly permuting security matches within industry and reopening date. The left panel uses the baseline regression; the right panel uses the common sample regression controlling for ES market exposure. The vertical lines indicate the coefficients under actual security matching.
\end{exhibitnotes}\end{minipage}\par\addvspace{7pt}

\par\medskip\noindent\begin{minipage}{\linewidth}
\subsection{Trading Venue Coverage and Sample Robustness}\label{app:D4}
\centering\setlength{\parskip}{0pt}
\captionof{table}{Robustness to venue coverage and sample selection}\label{tab:D4}
\begingroup\linespread{1.0}\fontsize{8}{9.0}\selectfont\setlength{\tabcolsep}{2.2pt}\renewcommand{\arraystretch}{1.0}
\begin{tabular}{@{}L{0.19\linewidth}*{5}{C{\dimexpr(\linewidth-0.19\linewidth-10\tabcolsep)/5\relax}}@{}}\toprule
\multicolumn{6}{@{}p{\linewidth}@{}}{\bfseries Panel A. Cumulative opening return regressions by venue}\\[0pt]
Venue & Perpetual return coefficient & Standard error & Stock-weeks & Stocks & Incr. within-\ensuremath{R^2} (\%)\\\midrule
MEXC & \regcoef{0}{696}{***} & \regse{0}{184} & 3,265 & 250 & \regstat{8}{84}\\
Bitget & \regcoef{0}{546}{***} & \regse{0}{174} & 2,931 & 220 & \regstat{6}{59}\\
Gate & \regcoef{0}{608}{***} & \regse{0}{175} & 1,998 & 178 & \regstat{8}{10}\\
KuCoin & \regcoef{0}{235}{***} & \regse{0}{071} & 1,790 & 121 & \regstat{1}{87}\\
Bitunix & \regcoef{0}{197}{**} & \regse{0}{091} & 1,442 & 132 & \regstat{1}{11}\\
Bybit & \regcoef{0}{338}{***} & \regse{0}{089} & 1,426 & 148 & \regstat{3}{87}\\
Phemex & \regcoef{0}{581}{***} & \regse{0}{141} & 1,231 & 90 & \regstat{9}{88}\\
BingX & \regcoef{1}{097}{**} & \regse{0}{309} & 908 & 205 & \regstat{16}{64}\\
HTX & \regcoef{0}{198}{*} & \regse{0}{115} & 674 & 95 & \regstat{1}{03}\\
XT & \regcoef{0}{192}{*} & \regse{0}{102} & 647 & 90 & \regstat{1}{02}\\
Binance & \regcoef{0}{897}{***} & \regse{0}{300} & 633 & 137 & \regstat{10}{47}\\
Ourbit & \regcoef{0}{333}{**} & \regse{0}{163} & 550 & 62 & \regstat{2}{19}\\
OKX & \regcoef{0}{500}{**} & \regse{0}{198} & 437 & 69 & \regstat{5}{95}\\
Helix & \regcoef{0}{304}{} & \regse{0}{341} & 279 & 72 & \regstat{0}{31}\\
Deepcoin & \regcoef{0}{599}{*} & \regse{0}{273} & 244 & 42 & \regstat{5}{00}\\
Extended & \regcoef{0}{592}{***} & \regse{0}{124} & 208 & 20 & \regstat{11}{82}\\
BloFin & \regcoef{0}{814}{*} & \regse{0}{420} & 139 & 27 & \regstat{9}{06}\\
edgeX & \regcoef{0}{416}{} & \regse{0}{381} & 123 & 19 & \regstat{2}{28}\\
Deribit & \regcoef{0}{728}{**} & \regse{0}{218} & 112 & 26 & \regstat{13}{94}\\
Aster & \regcoef{0}{955}{***} & \regse{0}{300} & 92 & 26 & \regstat{18}{99}\\
Crypto.com & \regcoef{1}{843}{***} & \regse{0}{484} & 48 & 24 & \regstat{15}{17}\\
Kraken & \textemdash{} & \textemdash{} & 23 & 6 & \textemdash{}\\
\midrule\end{tabular}\endgroup
\par\nointerlineskip
\begingroup\linespread{1.0}\fontsize{8}{9.0}\selectfont\setlength{\tabcolsep}{2.0pt}\renewcommand{\arraystretch}{1.0}
\begin{tabular}{@{}L{0.37\linewidth}*{6}{C{\dimexpr(\linewidth-0.37\linewidth-12\tabcolsep)/6\relax}}@{}}
\multicolumn{7}{@{}p{\linewidth}@{}}{\bfseries Panel B. Estimates with different levels of venue coverage}\\[0pt]
 & \multicolumn{6}{C{\dimexpr ((\linewidth-0.37\linewidth-12\tabcolsep)/6)*6+10\tabcolsep\relax}@{}}{Number of venues included, ranked by valid stock-weeks}\\
\cmidrule(lr){2-7}
Variable & 1 & 3 & 5 & 10 & 15 & 22\\\midrule
\rowlabel{Weekend perpetual return} & \regpair{\regcoef{0}{696}{***}}{\regse{0}{184}} & \regpair{\regcoef{0}{587}{***}}{\regse{0}{170}} & \regpair{\regcoef{0}{578}{***}}{\regse{0}{169}} & \regpair{\regcoef{0}{586}{***}}{\regse{0}{166}} & \regpair{\regcoef{0}{591}{***}}{\regse{0}{167}} & \regpair{\regcoef{0}{592}{***}}{\regse{0}{166}}\\
\addlinespace[0pt]
\rowlabel{Stock-weeks} & 3,265 & 3,973 & 3,990 & 4,033 & 4,053 & 4,060\\
\rowlabel{Stocks} & 250 & 305 & 305 & 313 & 318 & 319\\
\rowlabel{Within-\ensuremath{R^2}} & \regstat{0}{104} & \regstat{0}{079} & \regstat{0}{077} & \regstat{0}{080} & \regstat{0}{080} & \regstat{0}{080}\\
\rowlabel{Perpetual incr. within-\ensuremath{R^2} (\%)} & \regstat{8}{842} & \regstat{6}{698} & \regstat{6}{573} & \regstat{6}{932} & \regstat{6}{972} & \regstat{6}{994}\\
\midrule\end{tabular}\endgroup
\par\nointerlineskip
\begingroup\linespread{1.0}\fontsize{8}{9.0}\selectfont\setlength{\tabcolsep}{2.2pt}\renewcommand{\arraystretch}{1.0}
\begin{tabular}{@{}L{0.37\linewidth}*{2}{C{\dimexpr(\linewidth-0.37\linewidth-4\tabcolsep)/2\relax}}@{}}
\multicolumn{3}{@{}p{\linewidth}@{}}{\bfseries Panel C. Alternative sample coverage}\\[0pt]
Variable & Expanded sample & Expanded sample with historical stock characteristics\\\midrule
\rowlabel{Weekend perpetual return} & \regpair{\regcoef{0}{702}{***}}{\regse{0}{221}} & \regpair{\regcoef{0}{694}{***}}{\regse{0}{185}}\\
\addlinespace[0pt]
\rowlabel{Stock-weeks} & 7,489 & 7,456\\
\rowlabel{Stocks} & 421 & 421\\
\rowlabel{Within-\ensuremath{R^2}} & \regstat{0}{073} & \regstat{0}{103}\\
\rowlabel{Perpetual incr. within-\ensuremath{R^2} (\%)} & \regstat{7}{311} & \regstat{7}{041}\\
\midrule\end{tabular}\endgroup
\par\nointerlineskip
\begingroup\linespread{1.0}\fontsize{8}{9.0}\selectfont\setlength{\tabcolsep}{2.2pt}\renewcommand{\arraystretch}{1.0}
\begin{tabular}{@{}L{0.40\linewidth}*{2}{C{\dimexpr(\linewidth-0.40\linewidth-4\tabcolsep)/2\relax}}@{}}
\multicolumn{3}{@{}p{\linewidth}@{}}{\bfseries Panel D. Estimates excluding one venue at a time}\\[0pt]
 & Minimum coefficient & Maximum coefficient\\
Variable & Exclude Helix & Exclude Bitget\\\midrule
\rowlabel{Perpetual return coefficient} & \regpair{\regcoef{0}{697}{***}}{\regse{0}{219}} & \regpair{\regcoef{0}{742}{***}}{\regse{0}{230}}\\
\addlinespace[0pt]
\rowlabel{Stock-weeks} & 7,444 & 7,000\\
\rowlabel{Stocks} & 409 & 403\\
\rowlabel{Within-\ensuremath{R^2}} & \regstat{0}{073} & \regstat{0}{081}\\
\rowlabel{Perpetual incr. within-\ensuremath{R^2} (\%)} & \regstat{7}{270} & \regstat{8}{090}\\
\bottomrule\end{tabular}\endgroup
\begin{exhibitnotes}
All models include stock and reopening-date fixed effects. Panel A uses each venue's sample with a common return starting point and complete historical stock characteristics, requiring at least 30 stock-weeks, 3 stocks, and 3 reopening dates. Kraken does not meet the sample size requirement. Panel B ranks venues by valid stock-weeks and progressively expands venue coverage. Panel C compares the expanded sample with and without historical stock characteristics; stock returns begin at Friday 16:00 in both columns. Panel D excludes each of the 22 venues in turn from the expanded sample and reports the minimum and maximum perpetual return coefficients. Stock returns begin at Friday 16:00, and these regressions do not control for historical stock characteristics. Sample coverage differs across venues.\incrnote{}\regnote{}
\end{exhibitnotes}
\end{minipage}\par\medskip

\end{document}